\documentclass[aps,prl,reprint,groupedaddress]{revtex4-1}
\usepackage{amssymb}
\usepackage{amsmath}
\usepackage{graphicx}
\usepackage{epsfig} 
\usepackage{epic,eepic}
\usepackage{hyperref}
\usepackage{lipsum}
\usepackage[utf8]{inputenc}
\usepackage[vietnamese, english]{babel}

\newcommand{\removed}[1]{}

\begin{document}


\title{Majorana Zero-Modes in a Two-Dimensional $p$-Wave Superconductor}


\author{\foreignlanguage{vietnamese}{Võ Tiến Phong}$^1$}
\email{Phong.Vo@postgrad.manchester.ac.uk} 
\author{Niels R. Walet$^1$}
\email{Niels.Walet@manchester.ac.uk}
\author{Francisco Guinea$^{1,2}$}
\email{Francisco.Guinea@manchester.ac.uk}
\affiliation{$^1$School of Physics and Astronomy, University of Manchester, Manchester, M13 9PY, UK}
\affiliation{$^2$Imdea Nanoscience, Faraday 9, 28015 Madrid, Spain}

\date{\today}

\begin{abstract}
We analyze the formation of Majorana zero-modes at the edge of a two-dimensional topological superconductor. In particular, we study a time-reversal-invariant triplet phase that is likely to exist in doped Bi$_2$Se$_3$. Upon the introduction of an in-plane magnetic field to the superconductor, a gap is opened in the surface modes, which induces localized Majorana modes. The position of these modes can be simply manipulated by changing the orientation of the applied field, yielding novel methods for braiding these states with possible applications to topological quantum computation. 
\end{abstract}

\pacs{???}

\maketitle


Topological quantum computation is currently among the most interesting candidates for the realization of a universal quantum computer~\cite{BK16}. This approach provides a promising path to creating a robust qubit that can endure the necessary manipulations required in performing quantum logic~\cite{ SFN15}. Recent attempts at realizing such a qubit in condensed-matter platforms are motivated by the one-dimensional  Kitaev model~\cite{K01} with a topological insulating wire on which superconductivity  is induced by contact with an ordinary $s$-wave superconductor \cite{ORO10,LSS10}. Systems of such qubits are presently the subject of many investigations; see, for example, Ref.~\cite{BK16} and references contained therein. 

Many approaches to topological quantum computation are based on the creation and manipulation of massless Majorana states \cite{K01}. These arise as excitations in a two-dimensional system when a fermion is effectively split into two parts, with each part localized far away from the other in space. Since fermions are fundamental particles,  Majorana states are always generated in pairs. Such states are known to occur in topological superconductors \cite{HK10,LZ11,A12,LF12,AF15,AM16,KB16,SA16} and have been predicted to exist in the $\nu=5/2$ fractional quantum Hall effect \cite{RG00, FFMN07}. Because of their nontrivial half-fermion statistics, braiding, or exchanging, Majorana states is a non-Abelian process which takes place within the space of degenerate ground states. Quantum gates can be built by simply braiding Majorana states \cite{A12}. Currently, realistic schemes for  braiding  Majorana states require tri-wire junctions \cite{Aetal11}, and either electrostatic gates \cite{SCT11} or controllable magnetic fluxes \cite{HAHBB12}.

In this Letter, we propose and study a solid-state platform for the creation and manipulation of Majorana states, motivated by current research on  natural topological superconductors. Our proposed system is modeled on doped Bi$_2$Se$_3$, but the results we present here are more general, and can be extended to other superconductors and to the superfluid B phase of $^3$He. In what follows, we describe the general features of the our Majorana platform, and show by detailed calculations how these Majorana states arise. We then discuss a simple scheme for manipulating Majorana excitations on superconducting discs. Using a simple extension, we show that one can study our model in arbitrary geometries, providing further impetus for experimental realization.  This is described in full detail in the Appendix. 

\begin{figure}
\begin{center}
\includegraphics[width=7cm]{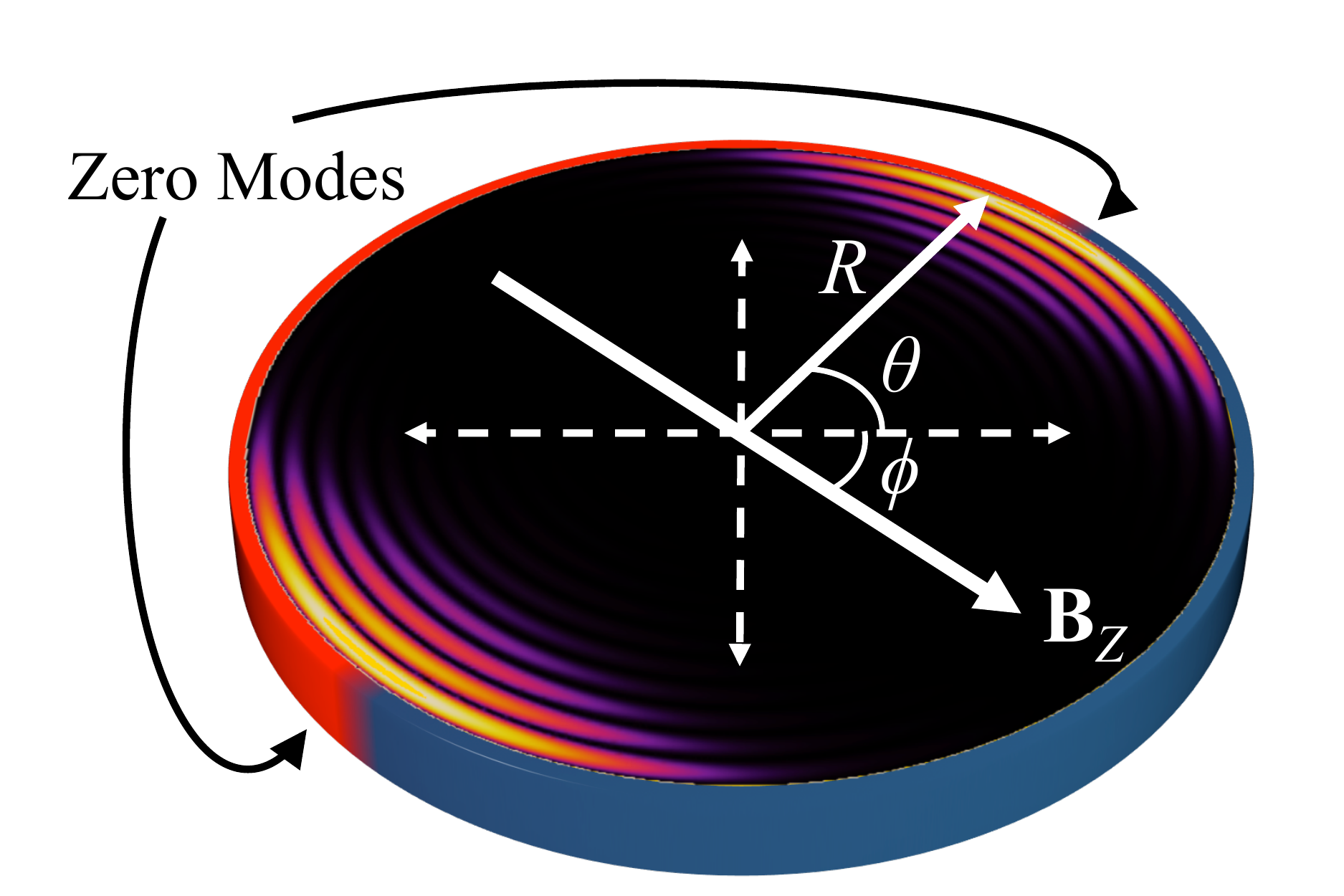}
\end{center}
\caption{Schematic representation of a superconducting disc in our setup. An in-plane magnetic field that gives rise to a Zeeman splitting is indicated by the arrow labeled $\mathbf{B}_Z.$ The color along the perimeter of the disc indicates relative sign of the superconducting gap at the edge. Majorana zero-modes are formed where the gap changes sign, as shown by the bright spots on the disc.}
\label{fig:sketch}
\end{figure}

We start with a thin disc of a topological superconductor material with thickness
smaller than the superconducting coherence length, as sketched in Fig.~\ref{fig:sketch}. In this limit, the surface states at the top and bottom surfaces
hybridize and develop a small gap.
The only remaining sub-gap states are quasi-one-dimensional gapless Majorana bands localized at the edges. If time-reversal symmetry (TRS) is not broken in the superconducting phase, there are two bands related by TRS inside the gap.
An in-plane magnetic field applied to the system, as shown in Fig.~\ref{fig:sketch}, breaks time-reversal symmetry. This field\removed{ thus} hybridizes the two sets of Andreev states, 
and opens a gap in the energy spectrum.  The sign and magnitude of the gap are determined by the normal of the field to the edge and is opposite where the field enters the disc to where the field exits the system. This leads to the formation of Majorana modes located \removed{in the middle of}inside the bulk energy gap. These Majorana edge modes are localized near the boundary of the disc, at the points where the field is parallel to the edge. 

Of the different order parameters that have been proposed to describe the superconducting phase of doped Bi$_2$Se$_3$~\cite{FB10,VKF16,CJG16}, we will consider  a time-reversal-invariant, odd-parity, triplet phase. Our analysis applies to similar Majorana modes appear in neutral fermionic superfluids, such as $^3$He~\cite{Letal13,Letal14,SIK16} (see also~\cite{TIM11}), as well as to gapped atomic Fermi superfluids~\cite{Ketal12}. In general, the two requirements for our proposed system  are: (i) (effective) two-dimensionality, and (ii) existence of gapless counter-propagating Andreev edge modes. The first condition requires the thickness 
of the system to be smaller than the superconducting coherence length. The second condition excludes a two-dimensional, gapped chiral $p_x \pm i p_y$ superconductor, as the corresponding edge modes flow in one direction only. Our results can be generalized to other topological superconducting phases, provided that the gap does not vanish on the Fermi surface, in particular to two-component time-reversal-invariant nematic phases~\cite{VKF16,CJG16}. Related artificial topological superconductors can  be created using the proximity effect~\cite{FK08,FKM13,bernardo_p-wave_2017}.


\begin{figure}
\begin{center}
\includegraphics[width=8cm]{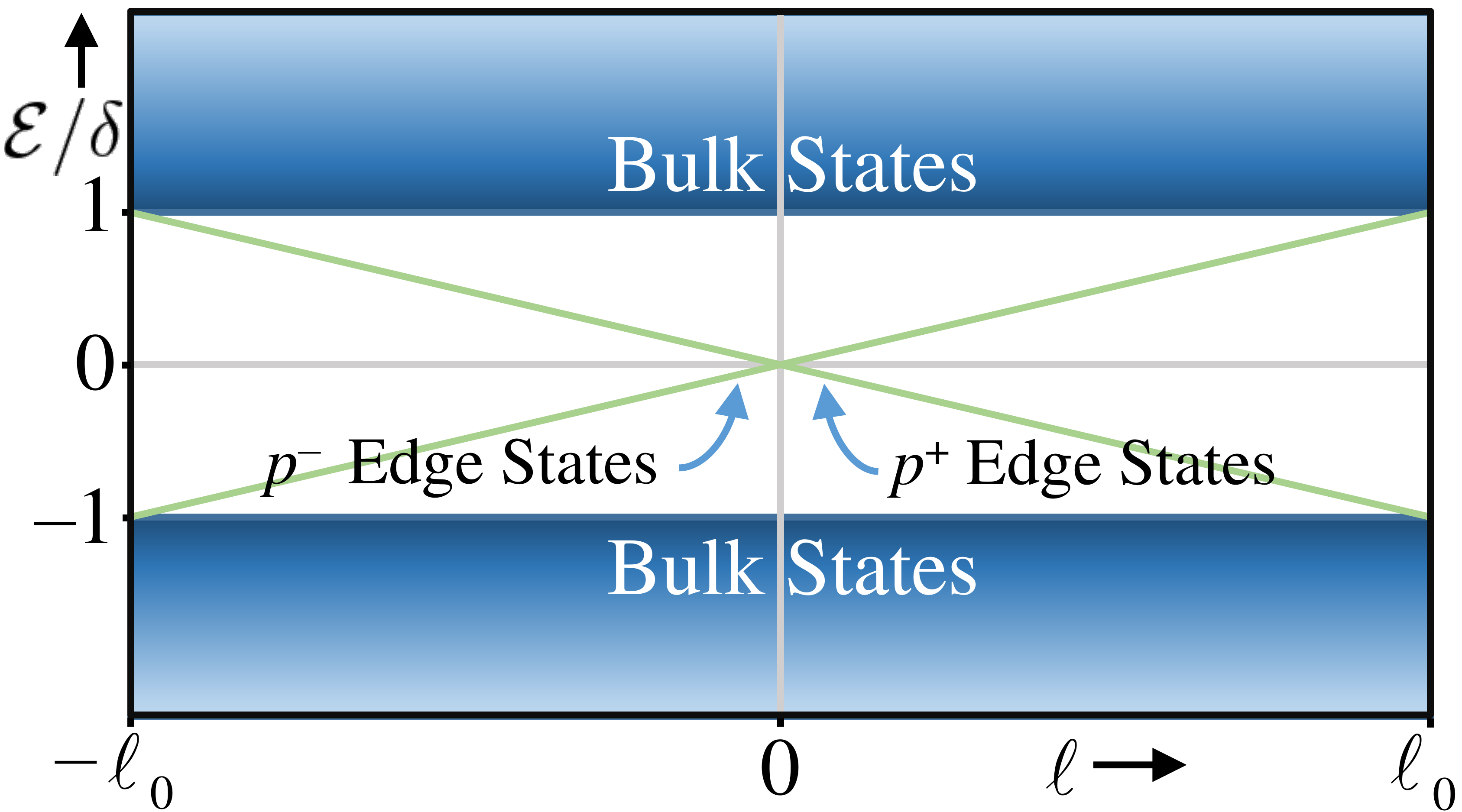}
\end{center}
\caption{Schematic of the spectrum of a time-reversal-invariant $p$-wave superconductor on a disc as a function of angular momentum $\ell$. Note that there are two branches of chiral edge modes with opposite chirality below the superconducting gap.}
\label{fig: band structure}
\end{figure}

We describe the system using a 
 two-orbital, $\mathbf{k}\cdot \mathbf{p}$ Hamiltonian proposed for the topological insulator Bi$_2$Se$_3$ ~\cite{zhang2009topological, PhysRevB.82.045122, PhysRevLett.108.107005}. 
 We consider quasi-two-dimensional systems, of thickness $d$ such that $k_{F z}^{-1} \ll d \lesssim \xi_z$, where $k_{F z}$ and $\xi_z$ are the Fermi wave-vector and the penetration length along the $z$-axis. The extraction of an effective low-energy two-dimensional Hamiltonian is described in the Appendix. We obtain
\begin{equation}
\mathcal{H}^\text{2D}(\mathbf{k}) = E_G \sigma_x   +v_F \sigma_z \left( k_x s_y-k_y s_x \right),
\end{equation}
where we denote by $\sigma_i$ and $s_i$ Pauli matrices that act on orbital space $\lbrace A, B \rbrace$ and spin space $\lbrace \uparrow,\downarrow \rbrace$, respectively, with $i = x,y,z$. We can induce superconductivity in this system via chemical doping. In the mean-field limit, we represent this by a Bogoliubov-de Gennes (BdG) Hamiltonian  
\begin{equation}
\mathcal{H}(\mathbf{k}) = [\mathcal{H}^\text{2D}(\mathbf{k})-\epsilon_F ]  \tau_z+ \Delta_\text{sc} \sigma_y s_z \tau_x,
\end{equation}
where the $\tau_i$ are Pauli matrices that act on the Nambu particle-hole space, $\epsilon_F$ is the chemical potential, and the coupling $\removed{\Delta_\text{sc} = }\Delta_\text{sc} \sigma_y s_z \tau_x$ describes a triplet $p$-wave, time-reversal-invariant superconductor.

For  Bi$_2$Se$_3$, and other systems of interest, the chemical potential lies far away from the valence band and the magnitude of the superconducting gap is small, i.e. $\bar{\epsilon}_F = \epsilon_F - E_G \lesssim 2E_G $ and $\Delta_\text{sc} \ll E_G$. In this limit, we can neglect the valence band, and project the BdG Hamiltonian onto the conduction band only. To lowest order in $\Delta_\text{sc}$, the Hamiltonian  reduces to two $2 \times 2$ Hamiltonians, $\mathcal{H}_{\pm}$. Thus, we write the projected Hamiltonian as
\begin{equation}
\label{eq: full Hamiltonian}
\mathcal{H}_0(\mathbf{k}) = \left( \begin{array}{ccc}
\mathcal{H}_+(\mathbf{k}) & 0 \\
0 & \mathcal{H}_-(\mathbf{k})   \end{array} \right),
\end{equation}
where $\mathcal{H}_\pm(\mathbf{k}) = \pm \left[ \frac{v_F^2|\mathbf{k}|^2}{2 E_G}-\bar{\epsilon}_F \right] \tau_z + \frac{2v_F\Delta_\text{sc}}{E_G} \left[k_y \tau_x + k_x \tau_y \right]$ represents two chiral superconductors with direction-dependent superconducting gaps at the Fermi surface,  $\Delta_\text{sc}^\text{Fermi} = 2 v_F k_F \Delta_\text{sc} e^{i\varphi}/E_G,$ where $\varphi = \arctan(k_y/k_x)$ and $k_F = \sqrt{2 E_G \bar{\epsilon}_F}/v_F.$ The measured superconducting gap is  $2\Delta_{\text{exp}} = 2 | \Delta_\text{sc}^\text{Fermi} | = 4 \Delta_\text{sc} v_F k_F / E_G$.  Each chiral superconductor has a single edge mode with a preferred direction ~\cite{RG00, PhysRevLett.102.187001}. The two $p$-wave branches are related by time-reversal symmetry, making the total Hamiltonian time-reversal invariant. Therefore, when solving for the eigenstates of $\mathcal{H}_0$ we restrict our attention to finding the states of $\mathcal{H}_+$, and determine  the eigenstates of $\mathcal{H}_-$ by reversing time.

We now study the edge modes on a disc of radius $R$ by diagonalizing $\mathcal{H}_0$ in that domain. In the following, we use dimensionless variables by expressing all energies in units of $\bar{\epsilon}_F$. Then, we can write
\begin{equation}
\label{eq: Hamiltonian for plus/minus sectors}
\mathcal{H}_\pm(\boldsymbol{\kappa}) = \left( \begin{array}{ccc}
\pm \left(|\boldsymbol{\kappa}|^2 - 1\right)& \sqrt{2\gamma} \left( -i\kappa_x+ \kappa_y  \right)  \\
\sqrt{2\gamma} \left( i\kappa_x+\kappa_y  \right) & \pm \left(-|\boldsymbol{\kappa}|^2 + 1 \right) \end{array} \right),
\end{equation}
where $\boldsymbol{\kappa} = \mathbf{k}/k_F,$ and $\gamma = \Delta_\text{exp}^2
 / 2 \bar{\epsilon}_F^2  \ll 1$ sets the relative scale of the superconducting gap. In our notation, the electron and hole components are ordered as $|\Psi\rangle = \left( \begin{array}{cccc}
\psi_\uparrow^{(e)} & \psi_\downarrow^{(h)} &- \psi_\uparrow^{(h)}& \psi_\downarrow^{(e)}  
\end{array} \right)^T$. The eigenfunctions of the Hamiltonian in Eq.~(\ref{eq: Hamiltonian for plus/minus sectors}) on a disc are two-component spinors,
where each component contains the product of radial Bessel functions with wave-vector $\kappa = |\boldsymbol{\kappa}|$ and  angular momentum $\ell$, as discussed in detail  in the appendix. The upper and lower components differ by one unit of angular momentum ~\cite{PhysRevB.69.184511}. We are interested in the case where $\kappa$ is complex, corresponding to edge modes with an exponentially decaying wavefunction. This occurs for energies, $E$, below the superconducting gap, i.e.  $E/\bar{\epsilon}_F = \mathcal{E} < \delta = \Delta_\text{sc}/\bar{\epsilon}_F.$ By imposing the boundary condition that the wavefunction vanishes on the edge, we can find the energy spectrum for the edge states. In the semiclassical limit where the system size is much larger than the Fermi wavelength $\lambda = k_F R \gg 1,$ and to first order in $\lambda^{-1}$, the spectra of edge modes for the  $p^+$ and $p^-$ superconductors are
\begin{equation}
\label{eq: edge mode energies}
\mathcal{E}_\pm^{\text{edge}} \approx \mp \left(\ell + \frac{1}{2} \right)\frac{\sqrt{2\gamma}}{\lambda}.
\end{equation}
We note that Eq.~(\ref{eq: edge mode energies}) is valid up to $|\ell| \approx \lambda\sqrt{1-\gamma/2}$; for larger angular momenta, the states lie outside the superconducting gap, and they are delocalized. Due to circular symmetry, the probability density of the edge states is uniform along the perimeter of the disc. The decay length towards the interior of the disc is $r_\ast \propto \xi=\bar{\epsilon}_F/(\Delta_\text{exp}k_F)  $. \removed{In order to have well-localized edge modes,} 
This length should be\removed{ much}
smaller than the disc's radius. Table~\ref{table:estimates} list estimates of the coherence length and of other parameters. Scalar disorder leads to pair breaking in $p$-wave superconductors, although the spin-orbit coupling reduces this effect~\cite{MF12}. The parameters in Table~\ref{table:estimates} use the critical temperatures taken from experiments~\cite{Hetal10}, which already include the effect of disorder. Finally,
it is important to note that scalar disorder does not mix the mid-gap states.

We now consider a symmetry-breaking perturbation that induces zero-mode Majorana states. We apply an in-plane magnetic field that breaks time-reversal symmetry and induces a superconducting gap that vanishes at boundary points where the field is tangent to the disc. We expect to find localized zero-modes at these points. The perturbation is of the form $\mathcal{H}_Z = \tau_Z \left( \mathcal{E}_Z \mathbf{n} \cdot \mathbf{s} \right),$ where $\mathbf{n}$ points in the direction of the field, $\mathcal{E}_Z = \mu_e B_Z / \bar{\epsilon}_F$ is the Zeeman coupling scaled by $\bar{\epsilon}_F,$  and $\mu_e$ is the magnetic moment of the electron, see  Fig.~\ref{fig:sketch}. The Zeeman field couples electrons of opposite spins from the different branches, and likewise, holes from different branches. Note that a parallel magnetic field weakly perturbs the superconducting phase~\cite{C62,C62b}. We work in the weak-field limit, where the Zeeman energy is much smaller than the superconducting gap.  Then, we can neglect the bulk states, and truncate the Hilbert space to the (unperturbed) edge states only. We project the full Hamiltonian $\mathcal{H} = \mathcal{H}_0 + \mathcal{H}_Z$ onto this basis and diagonalize the truncated Hamiltonian. We find two Majorana zero-modes separated in energy from the first excited quasi-particle states by an energy gap $\Delta_M$, which can be expressed as $\Delta_M / \bar{\epsilon}_F \approx \left(8 \gamma/\lambda^2 \times \mathcal{E}_Z^2 \right)^{1/4}$  for large $\lambda$. This implies stability of the zero-mode states as they are well-separated from the other edge modes. A second important scale is the splitting between the two Majorana states. Since the wavefunction at either side of the disc is approximately Gaussian, with angular width is $\langle \theta^2 \rangle \approx \sqrt{2 \gamma}/ ( {\cal E}_Z \lambda )$, this leads to a splitting which decays exponentially with $R$. As illustrated here, we can manipulate the gaps that determine the stability of the Majorana modes by simply adjusting the strength of the applied field  and the size of the disc.

\begin{table*}
\renewcommand*{\arraystretch}{1.2}
\begin{tabular}{lllll}
\hline \hline
\textbf{Input parameters:} & Radius &$R$& &$1\,\mu\mathrm{m}$ \\ 
& Magnetic field &$B_Z$ & &1 T \\ \hline

\textbf{Length and angular scales:} & Fermi wavelength &$k_F$ & &$10^{-1}\,\mathrm{\AA}^{-1}$ \\
& Electron density per layer&$\rho_L$ &${k_F^2}/{2 \pi}$ &$1.6 \times 10^{13}\,\mathrm{cm}^{-2}$ \\
& Coherence length &$\xi$ &${0.2 \times ( \hbar v_F ) / ( k_B T_c )}$ & $2 \times 10^3\,\mathrm{\AA}$\\
& Angular width of Majorana states &$1/b$ & $\sqrt{\Delta_\text{exp} /( \mu_e B_Z k_F R)}$& 0.32 radians\\ \hline

\textbf{Energy scales:} & Fermi energy &$\bar{\epsilon}_F$ & ${v_F^2 k_F^2}/{2 E_G}$ &$k_B \times 2300\,\mathrm{K} = 200\,\mathrm{meV}$ \\
& Critical temperature &$T_c$ & &3.8 K \\
& Superconducting gap  &$\Delta_\text{exp}$ & $1.76 \, k_BT_c$& $k_B \times 6.7\,\mathrm{K} = 0.6 \,\mathrm{meV}$ \\
& Quasi-particle gap &$\Delta_M$ &$\sqrt{{(2\mu_e B_Z\Delta_\text{exp} )}/{(k_F R )}}$& $k_B \times 0.3\, \mathrm{K} = 0.026 \, \mathrm{meV}$ \\
& Gap between Majorana modes &$\delta \epsilon_M$ &$\Delta_M  \exp(-2b^2)/\sqrt{2}$ &$ k_B\times 10^{-8} \, \mathrm{K} = 10^{-9} \, \mathrm{meV}$
\\
\hline \hline
\end{tabular}
\caption{Relevant parameters, analytical expressions, and numerical estimates of different quantities discussed in the text. Here, $k_B$ is Boltzmann's constant, and $\mu_e$ is the magnetic moment of the electron. We estimate these parameters based on experimental data available for Cu$_x$Bi$_2$Se$_3$ \cite{Hetal10,Wetal10, PhysRevB.84.054513, doi:10.1063/1.3200237, PhysRevB.90.094503, PhysRevB.83.224516}. Details can be found in the Appendix.} 
\label{table:estimates}
\end{table*}

\begin{figure}
\begin{center}
\includegraphics[width=8cm]{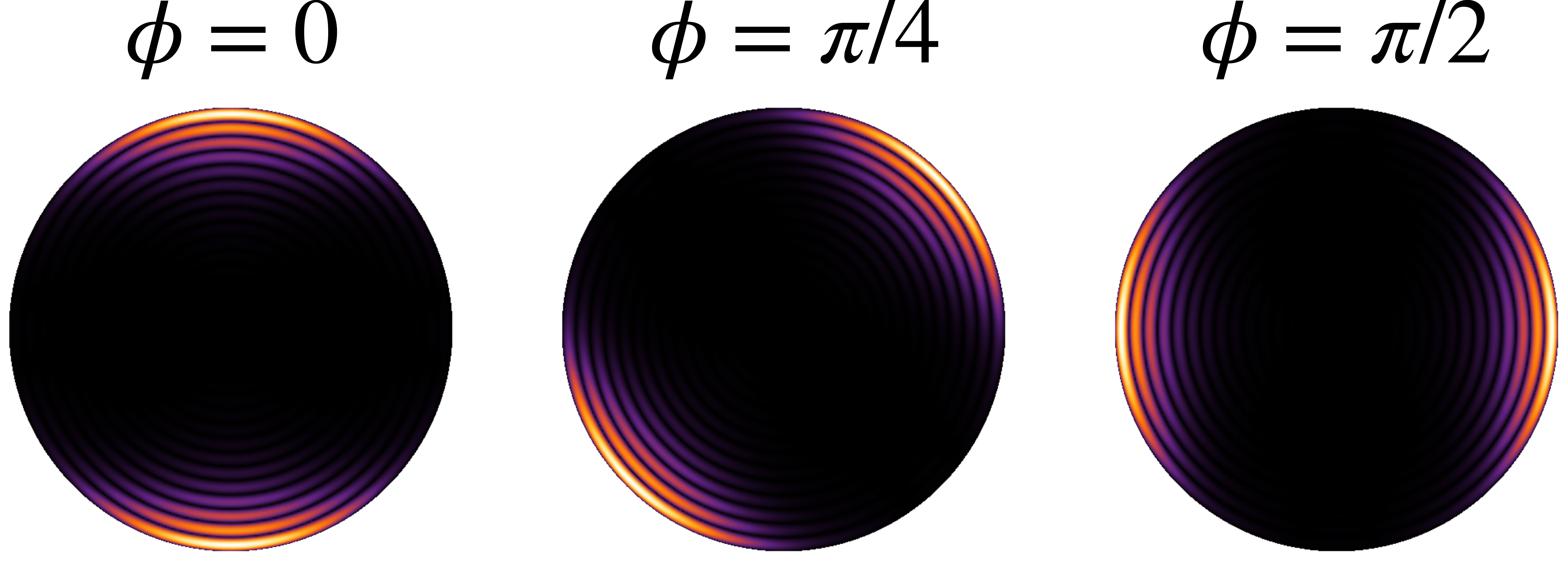}
\end{center}
\caption{Majorana zero-modes in the presence of a Zeeman field. We plot here the electron densities localized around points where the superconducting gap changes sign. We see that the zero-modes rotate as we rotate the in-plane field by changing $\phi.$ We use the following parameters: $\gamma 
	= {1}/{16},$ $\lambda =  k_F R = 50,$ and $\mathcal{E}_Z = {1}/{20}.$ 
}
\label{fig: wavefunction}
\end{figure}

\begin{figure}
\begin{center}
\includegraphics[width=8.5cm]{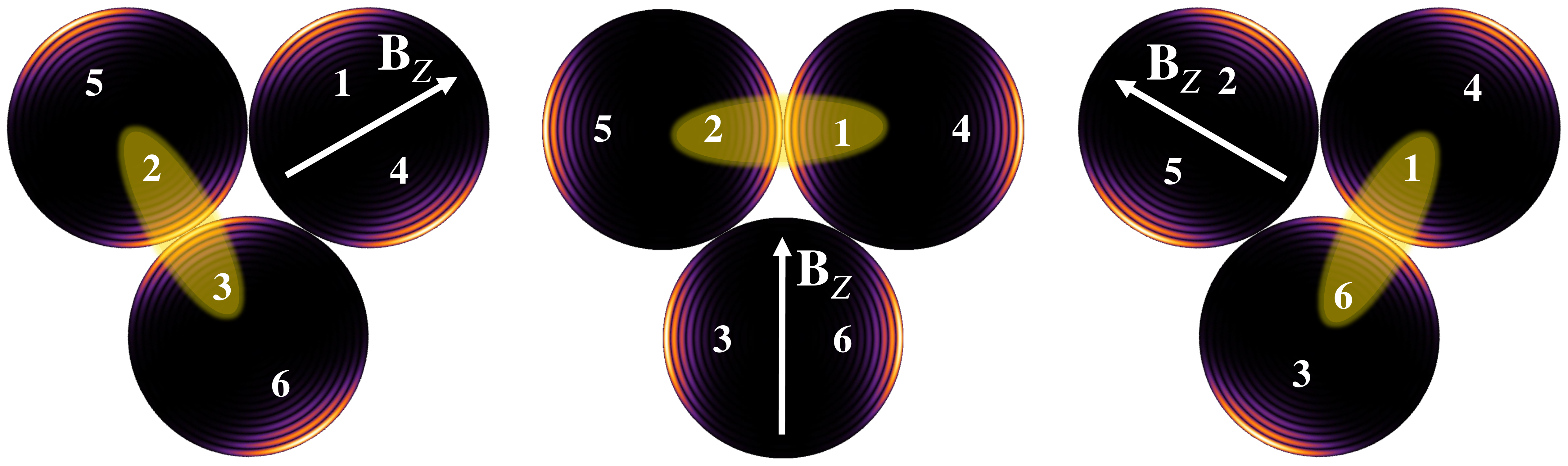}
\end{center}
\caption{Manipulation of Majorana modes by an in-plane magnetic field in an array of three topological superconductor quantum dots. The numbers label the Majorana states. Tunneling-assisted mixing between these states occurs when they are in nearby regions of the dots, as shown by the ellipses.}
\label{fig:qubit}
\end{figure}

As the edge modes are localized at boundary points where the superconducting gap changes sign, we can rotate these modes by simply rotating the Zeeman field, as shown Fig.~\ref{fig: wavefunction}. This gives us a simple way to perform particle exchanges for braiding and other purposes.  We now explore this possibility in a three-disc configuration.  
In the presence of tunneling junctions between discs, the Hamiltonian restricted to the  Majorana states is
\begin{align}
{\cal H}_M &= i \sum_{i,j} t_{i,j} ( \phi ) \gamma_i \gamma_j,
\label{hamilm}
\end{align}
where $\gamma_i$ and $\gamma_j$ are Majorana operators in neighboring discs, and $t_{i,j} ( \phi )$ is the hopping between those Majorana states. This term depends on the  orientation of the field, $\phi$. By rotating the magnetic field, states in different discs can be brought into contact, as illustrated in Fig.~\ref{fig:qubit}. For instance, the change of the exchange field which takes the left configuration into the center one exchanges states 3 and 1, while the change from the center configuration to the right one exchanges states 2 and 6. 
This scheme shows a simple way to manipulate the Majorana fermions. A complete braiding protocol is outside the scope of this work. An alternative proposal, based on a one-dimensonal ring of magnetic atoms on a superconductor has been recently discussed in~\cite{JNBY16} (note that the braiding scheme discussed there can be extended to our proposal, using overlapping dots). For other realizations,  see also~\cite{NDBY13,Netal14}.

The appeal of this approach is  that no tri-junctions are required, and only a rotation of the magnetic field is needed. These operations are carried out without the recourse to external gates or magnetic fluxes. If such fluxes and gates are added to this system, more ways to correlate the Majorana particles are induced, leading to new functionalities. Furthermore, more complex dot geometries can also be a platform for more ambitious engineering efforts. We explore some of these possibilities in the Appendix. 

It may
also be the case that the magnetic field is due to spontaneously polarized magnetic moments or to an additional chiral superconducting component of the order parameter~\cite{CJG16}. Then, quantum fluctuations of the field will lead to additional interactions between the Majorana states. Finally, a large array of quantum dots can serve as a platform for a surface code for topological quantum computation~\cite{BK01,VHF15,VF15}.

One question that needs to be addressed is whether we can perform the magnetic field rotation
without substantially mixing in low-lying fermionic excitations (so-called
quasiparticle poisoning \cite{SFN15}). As shown in detail in the Appendix, in the limit of weak coupling, we can adjust the separation in energy scales to be large enough to assure that this does not happen. There is a subtle interplay between
effective tunneling-supported mixing of the Majorana particles and quasiparticle poisoning that deserves further investigation.

In summary, we have analyzed the emergence of localized Majorana modes at the edges of finite two-dimensional topological superconductors. Though we have based our model on the time-reversal-invariant triplet superconducting phase that is likely to exist in doped Bi$_2$Se$_3,$ our results apply more generally to systems where the thickness is much smaller than the bulk coherence length. In particular, the triplet superconducting phase of doped Bi$_2$Se$_3$ bears qualitative resemblance to the B phase of superfluid $^3$He, and therefore, we expect our results to hold in that context as well. Furthermore, the analysis remains valid for other gapped superconducting phases with gapless edge modes, such as the nematic phase, also proposed for doped Bi$_2$Se$_3$. Similar Majorana states can also be expected in fermionic superfluids based on cold atoms. 

Our system provides an appealing alternative Majorana platform precisely because of its simplicity. We can control the stability of the Majorana states by widening the gap to the next excited states with a simple adjustment of the field strength. The positions of the Majorana states can be modified by changing the orientation of the applied magnetic field. Already with a single disc, we can exchange two Majorana particles by rotating the field by $\pi.$ In arrays with many quantum dots, the field can be used to modulate the interaction between Majorana states in different dots, and to exchange them, without requiring the existence of tri-junctions, electrostatic gates, or magnetic fluxes. These properties make the proposed model an interesting candidate for the experimental realization of a Majorana-based qubit for topological quantum computation.

\begin{acknowledgments}
We would like to thank R. Aguado, L. Chirolli, and P. San-Jose for helpful conversations. This work was supported by
funding from the European Union through the ERC Advanced
Grant NOVGRAPHENE through grant agreement
Nr. 290846, and from the European Commission under
the Graphene Flagship, contract CNECTICT-604391. VTP acknowledges financial support from the Marshall Aid Commemoration Commission. 
\end{acknowledgments}

\bibliography{arxivpaper,arxivpaper_v2}

\begin{thebibliography}{50}%
\makeatletter
\providecommand \@ifxundefined [1]{%
 \@ifx{#1\undefined}
}%
\providecommand \@ifnum [1]{%
 \ifnum #1\expandafter \@firstoftwo
 \else \expandafter \@secondoftwo
 \fi
}%
\providecommand \@ifx [1]{%
 \ifx #1\expandafter \@firstoftwo
 \else \expandafter \@secondoftwo
 \fi
}%
\providecommand \natexlab [1]{#1}%
\providecommand \enquote  [1]{``#1''}%
\providecommand \bibnamefont  [1]{#1}%
\providecommand \bibfnamefont [1]{#1}%
\providecommand \citenamefont [1]{#1}%
\providecommand \href@noop [0]{\@secondoftwo}%
\providecommand \href [0]{\begingroup \@sanitize@url \@href}%
\providecommand \@href[1]{\@@startlink{#1}\@@href}%
\providecommand \@@href[1]{\endgroup#1\@@endlink}%
\providecommand \@sanitize@url [0]{\catcode `\\12\catcode `\$12\catcode
  `\&12\catcode `\#12\catcode `\^12\catcode `\_12\catcode `\%12\relax}%
\providecommand \@@startlink[1]{}%
\providecommand \@@endlink[0]{}%
\providecommand \url  [0]{\begingroup\@sanitize@url \@url }%
\providecommand \@url [1]{\endgroup\@href {#1}{\urlprefix }}%
\providecommand \urlprefix  [0]{URL }%
\providecommand \Eprint [0]{\href }%
\providecommand \doibase [0]{http://dx.doi.org/}%
\providecommand \selectlanguage [0]{\@gobble}%
\providecommand \bibinfo  [0]{\@secondoftwo}%
\providecommand \bibfield  [0]{\@secondoftwo}%
\providecommand \translation [1]{[#1]}%
\providecommand \BibitemOpen [0]{}%
\providecommand \bibitemStop [0]{}%
\providecommand \bibitemNoStop [0]{.\EOS\space}%
\providecommand \EOS [0]{\spacefactor3000\relax}%
\providecommand \BibitemShut  [1]{\csname bibitem#1\endcsname}%
\let\auto@bib@innerbib\@empty
\bibitem [{\citenamefont {Beenakker}\ and\ \citenamefont
  {Kouwenhoven}(2016)}]{BK16}%
  \BibitemOpen
  \bibfield  {author} {\bibinfo {author} {\bibfnamefont {C.}~\bibnamefont
  {Beenakker}}\ and\ \bibinfo {author} {\bibfnamefont {L.}~\bibnamefont
  {Kouwenhoven}},\ }\href
  {http://www.nature.com/nphys/journal/v12/n7/full/nphys3778.html} {\bibfield
  {journal} {\bibinfo  {journal} {Nature Physics}\ }\textbf {\bibinfo {volume}
  {12}},\ \bibinfo {pages} {618} (\bibinfo {year} {2016})}\BibitemShut
  {NoStop}%
\bibitem [{\citenamefont {Sarma}\ \emph {et~al.}(2015)\citenamefont {Sarma},
  \citenamefont {Freedman},\ and\ \citenamefont {Nayak}}]{SFN15}%
  \BibitemOpen
  \bibfield  {author} {\bibinfo {author} {\bibfnamefont {S.~D.}\ \bibnamefont
  {Sarma}}, \bibinfo {author} {\bibfnamefont {M.}~\bibnamefont {Freedman}}, \
  and\ \bibinfo {author} {\bibfnamefont {C.}~\bibnamefont {Nayak}},\ }\href
  {http://dx.doi.org/10.1038/npjqi.2015.1} {\bibfield  {journal} {\bibinfo
  {journal} {NPJ Quantum Info.}\ }\textbf {\bibinfo {volume} {1}},\ \bibinfo
  {pages} {1083} (\bibinfo {year} {2015})}\BibitemShut {NoStop}%
\bibitem [{\citenamefont {Kitaev}(2001)}]{K01}%
  \BibitemOpen
  \bibfield  {author} {\bibinfo {author} {\bibfnamefont {A.~Y.}\ \bibnamefont
  {Kitaev}},\ }\href {http://stacks.iop.org/1063-7869/44/i=10S/a=S29}
  {\bibfield  {journal} {\bibinfo  {journal} {Physics-Uspekhi}\ }\textbf
  {\bibinfo {volume} {44}},\ \bibinfo {pages} {131} (\bibinfo {year}
  {2001})}\BibitemShut {NoStop}%
\bibitem [{\citenamefont {Oreg}\ \emph {et~al.}(2010)\citenamefont {Oreg},
  \citenamefont {Refael},\ and\ \citenamefont {von Oppen}}]{ORO10}%
  \BibitemOpen
  \bibfield  {author} {\bibinfo {author} {\bibfnamefont {Y.}~\bibnamefont
  {Oreg}}, \bibinfo {author} {\bibfnamefont {G.}~\bibnamefont {Refael}}, \ and\
  \bibinfo {author} {\bibfnamefont {F.}~\bibnamefont {von Oppen}},\ }\href
  {\doibase 10.1103/PhysRevLett.105.177002} {\bibfield  {journal} {\bibinfo
  {journal} {Phys. Rev. Lett.}\ }\textbf {\bibinfo {volume} {105}},\ \bibinfo
  {pages} {177002} (\bibinfo {year} {2010})}\BibitemShut {NoStop}%
\bibitem [{\citenamefont {Lutchyn}\ \emph {et~al.}(2010)\citenamefont
  {Lutchyn}, \citenamefont {Sau},\ and\ \citenamefont {Das~Sarma}}]{LSS10}%
  \BibitemOpen
  \bibfield  {author} {\bibinfo {author} {\bibfnamefont {R.~M.}\ \bibnamefont
  {Lutchyn}}, \bibinfo {author} {\bibfnamefont {J.~D.}\ \bibnamefont {Sau}}, \
  and\ \bibinfo {author} {\bibfnamefont {S.}~\bibnamefont {Das~Sarma}},\ }\href
  {\doibase 10.1103/PhysRevLett.105.077001} {\bibfield  {journal} {\bibinfo
  {journal} {Phys. Rev. Lett.}\ }\textbf {\bibinfo {volume} {105}},\ \bibinfo
  {pages} {077001} (\bibinfo {year} {2010})}\BibitemShut {NoStop}%
\bibitem [{\citenamefont {Hasan}\ and\ \citenamefont {Kane}(2010)}]{HK10}%
  \BibitemOpen
  \bibfield  {author} {\bibinfo {author} {\bibfnamefont {M.~Z.}\ \bibnamefont
  {Hasan}}\ and\ \bibinfo {author} {\bibfnamefont {C.~L.}\ \bibnamefont
  {Kane}},\ }\href {\doibase 10.1103/RevModPhys.82.3045} {\bibfield  {journal}
  {\bibinfo  {journal} {Rev. Mod. Phys.}\ }\textbf {\bibinfo {volume} {82}},\
  \bibinfo {pages} {3045} (\bibinfo {year} {2010})}\BibitemShut {NoStop}%
\bibitem [{\citenamefont {Qi}\ and\ \citenamefont {Zhang}(2011)}]{LZ11}%
  \BibitemOpen
  \bibfield  {author} {\bibinfo {author} {\bibfnamefont {X.-L.}\ \bibnamefont
  {Qi}}\ and\ \bibinfo {author} {\bibfnamefont {S.-C.}\ \bibnamefont {Zhang}},\
  }\href {\doibase 10.1103/RevModPhys.83.1057} {\bibfield  {journal} {\bibinfo
  {journal} {Rev. Mod. Phys.}\ }\textbf {\bibinfo {volume} {83}},\ \bibinfo
  {pages} {1057} (\bibinfo {year} {2011})}\BibitemShut {NoStop}%
\bibitem [{\citenamefont {Alicea}(2012)}]{A12}%
  \BibitemOpen
  \bibfield  {author} {\bibinfo {author} {\bibfnamefont {J.}~\bibnamefont
  {Alicea}},\ }\href {http://stacks.iop.org/0034-4885/75/i=7/a=076501}
  {\bibfield  {journal} {\bibinfo  {journal} {Rep. Prog. Phys.}\ }\textbf
  {\bibinfo {volume} {75}},\ \bibinfo {pages} {076501} (\bibinfo {year}
  {2012})}\BibitemShut {NoStop}%
\bibitem [{\citenamefont {Leijnse}\ and\ \citenamefont
  {Flensberg}(2012)}]{LF12}%
  \BibitemOpen
  \bibfield  {author} {\bibinfo {author} {\bibfnamefont {M.}~\bibnamefont
  {Leijnse}}\ and\ \bibinfo {author} {\bibfnamefont {K.}~\bibnamefont
  {Flensberg}},\ }\href {http://stacks.iop.org/0268-1242/27/i=12/a=124003}
  {\bibfield  {journal} {\bibinfo  {journal} {Semicond. Sci. Technol.}\
  }\textbf {\bibinfo {volume} {27}},\ \bibinfo {pages} {124003} (\bibinfo
  {year} {2012})}\BibitemShut {NoStop}%
\bibitem [{\citenamefont {Ando}\ and\ \citenamefont {Fu}(2015)}]{AF15}%
  \BibitemOpen
  \bibfield  {author} {\bibinfo {author} {\bibfnamefont {Y.}~\bibnamefont
  {Ando}}\ and\ \bibinfo {author} {\bibfnamefont {L.}~\bibnamefont {Fu}},\
  }\href {\doibase 10.1146/annurev-conmatphys-031214-014501} {\bibfield
  {journal} {\bibinfo  {journal} {Ann. Rev. Condens. Matter Phys.}\ }\textbf
  {\bibinfo {volume} {6}},\ \bibinfo {pages} {361} (\bibinfo {year}
  {2015})}\BibitemShut {NoStop}%
\bibitem [{\citenamefont {Aasen}\ \emph {et~al.}(2016)\citenamefont {Aasen},
  \citenamefont {Hell}, \citenamefont {Mishmash}, \citenamefont {Higginbotham},
  \citenamefont {Danon}, \citenamefont {Leijnse}, \citenamefont {Jespersen},
  \citenamefont {Folk}, \citenamefont {Marcus}, \citenamefont {Flensberg},\
  and\ \citenamefont {Alicea}}]{AM16}%
  \BibitemOpen
  \bibfield  {author} {\bibinfo {author} {\bibfnamefont {D.}~\bibnamefont
  {Aasen}}, \bibinfo {author} {\bibfnamefont {M.}~\bibnamefont {Hell}},
  \bibinfo {author} {\bibfnamefont {R.~V.}\ \bibnamefont {Mishmash}}, \bibinfo
  {author} {\bibfnamefont {A.}~\bibnamefont {Higginbotham}}, \bibinfo {author}
  {\bibfnamefont {J.}~\bibnamefont {Danon}}, \bibinfo {author} {\bibfnamefont
  {M.}~\bibnamefont {Leijnse}}, \bibinfo {author} {\bibfnamefont {T.~S.}\
  \bibnamefont {Jespersen}}, \bibinfo {author} {\bibfnamefont {J.~A.}\
  \bibnamefont {Folk}}, \bibinfo {author} {\bibfnamefont {C.~M.}\ \bibnamefont
  {Marcus}}, \bibinfo {author} {\bibfnamefont {K.}~\bibnamefont {Flensberg}}, \
  and\ \bibinfo {author} {\bibfnamefont {J.}~\bibnamefont {Alicea}},\ }\href
  {\doibase 10.1103/PhysRevX.6.031016} {\bibfield  {journal} {\bibinfo
  {journal} {Phys. Rev. X}\ }\textbf {\bibinfo {volume} {6}},\ \bibinfo {pages}
  {031016} (\bibinfo {year} {2016})}\BibitemShut {NoStop}%
\bibitem [{\citenamefont {Kallin}\ and\ \citenamefont
  {Berlinsky}(2016)}]{KB16}%
  \BibitemOpen
  \bibfield  {author} {\bibinfo {author} {\bibfnamefont {C.}~\bibnamefont
  {Kallin}}\ and\ \bibinfo {author} {\bibfnamefont {J.}~\bibnamefont
  {Berlinsky}},\ }\href {http://stacks.iop.org/0034-4885/79/i=5/a=054502}
  {\bibfield  {journal} {\bibinfo  {journal} {Rep. Prog. Phys.}\ }\textbf
  {\bibinfo {volume} {79}},\ \bibinfo {pages} {054502} (\bibinfo {year}
  {2016})}\BibitemShut {NoStop}%
\bibitem [{\citenamefont {Sato}\ and\ \citenamefont {Ando}(2016)}]{SA16}%
  \BibitemOpen
  \bibfield  {author} {\bibinfo {author} {\bibfnamefont {M.}~\bibnamefont
  {Sato}}\ and\ \bibinfo {author} {\bibfnamefont {Y.}~\bibnamefont {Ando}},\
  }\href@noop {} {\enquote {\bibinfo {title} {Topological superconductors},}\ }
  (\bibinfo {year} {2016}),\ \bibinfo {note} {preprint},\ \Eprint
  {http://arxiv.org/abs/1608.03395} {arXiv:1608.03395} \BibitemShut {NoStop}%
\bibitem [{\citenamefont {Read}\ and\ \citenamefont {Green}(2000)}]{RG00}%
  \BibitemOpen
  \bibfield  {author} {\bibinfo {author} {\bibfnamefont {N.}~\bibnamefont
  {Read}}\ and\ \bibinfo {author} {\bibfnamefont {D.}~\bibnamefont {Green}},\
  }\href {\doibase 10.1103/PhysRevB.61.10267} {\bibfield  {journal} {\bibinfo
  {journal} {Phys. Rev. B}\ }\textbf {\bibinfo {volume} {61}},\ \bibinfo
  {pages} {10267} (\bibinfo {year} {2000})}\BibitemShut {NoStop}%
\bibitem [{\citenamefont {Fendley}\ \emph {et~al.}(2007)\citenamefont
  {Fendley}, \citenamefont {Fisher},\ and\ \citenamefont {Nayak}}]{FFMN07}%
  \BibitemOpen
  \bibfield  {author} {\bibinfo {author} {\bibfnamefont {P.}~\bibnamefont
  {Fendley}}, \bibinfo {author} {\bibfnamefont {M.~P.~A.}\ \bibnamefont
  {Fisher}}, \ and\ \bibinfo {author} {\bibfnamefont {C.}~\bibnamefont
  {Nayak}},\ }\href {\doibase 10.1103/PhysRevB.75.045317} {\bibfield  {journal}
  {\bibinfo  {journal} {Phys. Rev. B}\ }\textbf {\bibinfo {volume} {75}},\
  \bibinfo {pages} {045317} (\bibinfo {year} {2007})}\BibitemShut {NoStop}%
\bibitem [{\citenamefont {Alicea}\ \emph {et~al.}(2011)\citenamefont {Alicea},
  \citenamefont {Oreg}, \citenamefont {Refael}, \citenamefont {von Oppen},\
  and\ \citenamefont {Fisher}}]{Aetal11}%
  \BibitemOpen
  \bibfield  {author} {\bibinfo {author} {\bibfnamefont {J.}~\bibnamefont
  {Alicea}}, \bibinfo {author} {\bibfnamefont {Y.}~\bibnamefont {Oreg}},
  \bibinfo {author} {\bibfnamefont {G.}~\bibnamefont {Refael}}, \bibinfo
  {author} {\bibfnamefont {F.}~\bibnamefont {von Oppen}}, \ and\ \bibinfo
  {author} {\bibfnamefont {M.~P.}\ \bibnamefont {Fisher}},\ }\href
  {http://www.nature.com/nphys/journal/v7/n5/abs/nphys1915.html} {\bibfield
  {journal} {\bibinfo  {journal} {Nature Physics}\ }\textbf {\bibinfo {volume}
  {7}},\ \bibinfo {pages} {412} (\bibinfo {year} {2011})}\BibitemShut {NoStop}%
\bibitem [{\citenamefont {Sau}\ \emph {et~al.}(2011)\citenamefont {Sau},
  \citenamefont {Clarke},\ and\ \citenamefont {Tewari}}]{SCT11}%
  \BibitemOpen
  \bibfield  {author} {\bibinfo {author} {\bibfnamefont {J.~D.}\ \bibnamefont
  {Sau}}, \bibinfo {author} {\bibfnamefont {D.~J.}\ \bibnamefont {Clarke}}, \
  and\ \bibinfo {author} {\bibfnamefont {S.}~\bibnamefont {Tewari}},\ }\href
  {\doibase 10.1103/PhysRevB.84.094505} {\bibfield  {journal} {\bibinfo
  {journal} {Phys. Rev. B}\ }\textbf {\bibinfo {volume} {84}},\ \bibinfo
  {pages} {094505} (\bibinfo {year} {2011})}\BibitemShut {NoStop}%
\bibitem [{\citenamefont {van Heck}\ \emph {et~al.}(2012)\citenamefont {van
  Heck}, \citenamefont {Akhmerov}, \citenamefont {Hassler}, \citenamefont
  {Burrello},\ and\ \citenamefont {Beenakker}}]{HAHBB12}%
  \BibitemOpen
  \bibfield  {author} {\bibinfo {author} {\bibfnamefont {B.}~\bibnamefont {van
  Heck}}, \bibinfo {author} {\bibfnamefont {A.~R.}\ \bibnamefont {Akhmerov}},
  \bibinfo {author} {\bibfnamefont {F.}~\bibnamefont {Hassler}}, \bibinfo
  {author} {\bibfnamefont {M.}~\bibnamefont {Burrello}}, \ and\ \bibinfo
  {author} {\bibfnamefont {C.~W.~J.}\ \bibnamefont {Beenakker}},\ }\href
  {http://stacks.iop.org/1367-2630/14/i=3/a=035019} {\bibfield  {journal}
  {\bibinfo  {journal} {New J. Phys.}\ }\textbf {\bibinfo {volume} {14}},\
  \bibinfo {pages} {035019} (\bibinfo {year} {2012})}\BibitemShut {NoStop}%
\bibitem [{\citenamefont {Fu}\ and\ \citenamefont {Berg}(2010)}]{FB10}%
  \BibitemOpen
  \bibfield  {author} {\bibinfo {author} {\bibfnamefont {L.}~\bibnamefont
  {Fu}}\ and\ \bibinfo {author} {\bibfnamefont {E.}~\bibnamefont {Berg}},\
  }\href {\doibase 10.1103/PhysRevLett.105.097001} {\bibfield  {journal}
  {\bibinfo  {journal} {Phys. Rev. Lett.}\ }\textbf {\bibinfo {volume} {105}},\
  \bibinfo {pages} {097001} (\bibinfo {year} {2010})}\BibitemShut {NoStop}%
\bibitem [{\citenamefont {Venderbos}\ \emph {et~al.}(2016)\citenamefont
  {Venderbos}, \citenamefont {Kozii},\ and\ \citenamefont {Fu}}]{VKF16}%
  \BibitemOpen
  \bibfield  {author} {\bibinfo {author} {\bibfnamefont {J.~W.~F.}\
  \bibnamefont {Venderbos}}, \bibinfo {author} {\bibfnamefont {V.}~\bibnamefont
  {Kozii}}, \ and\ \bibinfo {author} {\bibfnamefont {L.}~\bibnamefont {Fu}},\
  }\href {\doibase 10.1103/PhysRevB.94.180504} {\bibfield  {journal} {\bibinfo
  {journal} {Phys. Rev. B}\ }\textbf {\bibinfo {volume} {94}},\ \bibinfo
  {pages} {180504} (\bibinfo {year} {2016})}\BibitemShut {NoStop}%
\bibitem [{\citenamefont {Chirolli}\ \emph {et~al.}(2016)\citenamefont
  {Chirolli}, \citenamefont {{de Juan}},\ and\ \citenamefont {Guinea}}]{CJG16}%
  \BibitemOpen
  \bibfield  {author} {\bibinfo {author} {\bibfnamefont {L.}~\bibnamefont
  {Chirolli}}, \bibinfo {author} {\bibfnamefont {F.}~\bibnamefont {{de Juan}}},
  \ and\ \bibinfo {author} {\bibfnamefont {F.}~\bibnamefont {Guinea}},\ }\href
  {https://arxiv.org/abs/1611.02173} {\enquote {\bibinfo {title} {Time-reversal
  symmetry breaking superconductivity in dirac materials: application to
  $\mathrm{Nb_xBi_2Se_3}$},}\ } (\bibinfo {year} {2016}),\ \bibinfo {note}
  {preprint},\ \Eprint {http://arxiv.org/abs/1611.02173} {arXiv:1611.02173}
  \BibitemShut {NoStop}%
\bibitem [{\citenamefont {Levitin}\ \emph {et~al.}(2013)\citenamefont
  {Levitin}, \citenamefont {Bennett}, \citenamefont {Casey}, \citenamefont
  {Cowan}, \citenamefont {Saunders}, \citenamefont {Drung}, \citenamefont
  {Schurig},\ and\ \citenamefont {Parpia}}]{Letal13}%
  \BibitemOpen
  \bibfield  {author} {\bibinfo {author} {\bibfnamefont {L.~V.}\ \bibnamefont
  {Levitin}}, \bibinfo {author} {\bibfnamefont {R.~G.}\ \bibnamefont
  {Bennett}}, \bibinfo {author} {\bibfnamefont {A.}~\bibnamefont {Casey}},
  \bibinfo {author} {\bibfnamefont {B.}~\bibnamefont {Cowan}}, \bibinfo
  {author} {\bibfnamefont {J.}~\bibnamefont {Saunders}}, \bibinfo {author}
  {\bibfnamefont {D.}~\bibnamefont {Drung}}, \bibinfo {author} {\bibfnamefont
  {T.}~\bibnamefont {Schurig}}, \ and\ \bibinfo {author} {\bibfnamefont
  {J.~M.}\ \bibnamefont {Parpia}},\ }\href {\doibase 10.1126/science.1233621}
  {\bibfield  {journal} {\bibinfo  {journal} {Science}\ }\textbf {\bibinfo
  {volume} {340}},\ \bibinfo {pages} {841} (\bibinfo {year}
  {2013})}\BibitemShut {NoStop}%
\bibitem [{\citenamefont {Levitin}\ \emph {et~al.}(2014)\citenamefont
  {Levitin}, \citenamefont {Bennett}, \citenamefont {Casey}, \citenamefont
  {Cowan}, \citenamefont {Saunders}, \citenamefont {Drung}, \citenamefont
  {Schurig}, \citenamefont {Parpia}, \citenamefont {Ilic},\ and\ \citenamefont
  {Zhelev}}]{Letal14}%
  \BibitemOpen
  \bibfield  {author} {\bibinfo {author} {\bibfnamefont {L.~V.}\ \bibnamefont
  {Levitin}}, \bibinfo {author} {\bibfnamefont {R.~G.}\ \bibnamefont
  {Bennett}}, \bibinfo {author} {\bibfnamefont {A.}~\bibnamefont {Casey}},
  \bibinfo {author} {\bibfnamefont {B.}~\bibnamefont {Cowan}}, \bibinfo
  {author} {\bibfnamefont {J.}~\bibnamefont {Saunders}}, \bibinfo {author}
  {\bibfnamefont {D.}~\bibnamefont {Drung}}, \bibinfo {author} {\bibfnamefont
  {T.}~\bibnamefont {Schurig}}, \bibinfo {author} {\bibfnamefont {J.~M.}\
  \bibnamefont {Parpia}}, \bibinfo {author} {\bibfnamefont {B.}~\bibnamefont
  {Ilic}}, \ and\ \bibinfo {author} {\bibfnamefont {N.}~\bibnamefont
  {Zhelev}},\ }\href {\doibase 10.1007/s10909-014-1145-1} {\bibfield  {journal}
  {\bibinfo  {journal} {J. Low Temp. Phys.}\ }\textbf {\bibinfo {volume}
  {175}},\ \bibinfo {pages} {667} (\bibinfo {year} {2014})}\BibitemShut
  {NoStop}%
\bibitem [{\citenamefont {Saitoh}\ \emph {et~al.}(2016)\citenamefont {Saitoh},
  \citenamefont {Ikegami},\ and\ \citenamefont {Kono}}]{SIK16}%
  \BibitemOpen
  \bibfield  {author} {\bibinfo {author} {\bibfnamefont {M.}~\bibnamefont
  {Saitoh}}, \bibinfo {author} {\bibfnamefont {H.}~\bibnamefont {Ikegami}}, \
  and\ \bibinfo {author} {\bibfnamefont {K.}~\bibnamefont {Kono}},\ }\href
  {\doibase 10.1103/PhysRevLett.117.205302} {\bibfield  {journal} {\bibinfo
  {journal} {Phys. Rev. Lett.}\ }\textbf {\bibinfo {volume} {117}},\ \bibinfo
  {pages} {205302} (\bibinfo {year} {2016})}\BibitemShut {NoStop}%
\bibitem [{\citenamefont {Tsutsumi}\ \emph {et~al.}(2011)\citenamefont
  {Tsutsumi}, \citenamefont {Ichioka},\ and\ \citenamefont {Machida}}]{TIM11}%
  \BibitemOpen
  \bibfield  {author} {\bibinfo {author} {\bibfnamefont {Y.}~\bibnamefont
  {Tsutsumi}}, \bibinfo {author} {\bibfnamefont {M.}~\bibnamefont {Ichioka}}, \
  and\ \bibinfo {author} {\bibfnamefont {K.}~\bibnamefont {Machida}},\ }\href
  {\doibase 10.1103/PhysRevB.83.094510} {\bibfield  {journal} {\bibinfo
  {journal} {Phys. Rev. B}\ }\textbf {\bibinfo {volume} {83}},\ \bibinfo
  {pages} {094510} (\bibinfo {year} {2011})}\BibitemShut {NoStop}%
\bibitem [{\citenamefont {Ku}\ \emph {et~al.}(2012)\citenamefont {Ku},
  \citenamefont {Sommer}, \citenamefont {Cheuk},\ and\ \citenamefont
  {Zwierlein}}]{Ketal12}%
  \BibitemOpen
  \bibfield  {author} {\bibinfo {author} {\bibfnamefont {M.~J.~H.}\
  \bibnamefont {Ku}}, \bibinfo {author} {\bibfnamefont {A.~T.}\ \bibnamefont
  {Sommer}}, \bibinfo {author} {\bibfnamefont {L.~W.}\ \bibnamefont {Cheuk}}, \
  and\ \bibinfo {author} {\bibfnamefont {M.~W.}\ \bibnamefont {Zwierlein}},\
  }\href {\doibase 10.1126/science.1214987} {\bibfield  {journal} {\bibinfo
  {journal} {Science}\ }\textbf {\bibinfo {volume} {335}},\ \bibinfo {pages}
  {563} (\bibinfo {year} {2012})}\BibitemShut {NoStop}%
\bibitem [{\citenamefont {Fu}\ and\ \citenamefont {Kane}(2008)}]{FK08}%
  \BibitemOpen
  \bibfield  {author} {\bibinfo {author} {\bibfnamefont {L.}~\bibnamefont
  {Fu}}\ and\ \bibinfo {author} {\bibfnamefont {C.~L.}\ \bibnamefont {Kane}},\
  }\href {\doibase 10.1103/PhysRevLett.100.096407} {\bibfield  {journal}
  {\bibinfo  {journal} {Phys. Rev. Lett.}\ }\textbf {\bibinfo {volume} {100}},\
  \bibinfo {pages} {096407} (\bibinfo {year} {2008})}\BibitemShut {NoStop}%
\bibitem [{\citenamefont {Zhang}\ \emph {et~al.}(2013)\citenamefont {Zhang},
  \citenamefont {Kane},\ and\ \citenamefont {Mele}}]{FKM13}%
  \BibitemOpen
  \bibfield  {author} {\bibinfo {author} {\bibfnamefont {F.}~\bibnamefont
  {Zhang}}, \bibinfo {author} {\bibfnamefont {C.~L.}\ \bibnamefont {Kane}}, \
  and\ \bibinfo {author} {\bibfnamefont {E.~J.}\ \bibnamefont {Mele}},\ }\href
  {\doibase 10.1103/PhysRevLett.111.056402} {\bibfield  {journal} {\bibinfo
  {journal} {Phys. Rev. Lett.}\ }\textbf {\bibinfo {volume} {111}},\ \bibinfo
  {pages} {056402} (\bibinfo {year} {2013})}\BibitemShut {NoStop}%
\bibitem [{\citenamefont {Bernardo}\ \emph {et~al.}(2017)\citenamefont
  {Bernardo}, \citenamefont {Millo}, \citenamefont {Barbone}, \citenamefont
  {Alpern}, \citenamefont {Kalcheim}, \citenamefont {Sassi}, \citenamefont
  {Ott}, \citenamefont {Fazio}, \citenamefont {Yoon}, \citenamefont {Amado},
  \citenamefont {Ferrari}, \citenamefont {Linder},\ and\ \citenamefont
  {Robinson}}]{bernardo_p-wave_2017}%
  \BibitemOpen
  \bibfield  {author} {\bibinfo {author} {\bibfnamefont {A.~D.}\ \bibnamefont
  {Bernardo}}, \bibinfo {author} {\bibfnamefont {O.}~\bibnamefont {Millo}},
  \bibinfo {author} {\bibfnamefont {M.}~\bibnamefont {Barbone}}, \bibinfo
  {author} {\bibfnamefont {H.}~\bibnamefont {Alpern}}, \bibinfo {author}
  {\bibfnamefont {Y.}~\bibnamefont {Kalcheim}}, \bibinfo {author}
  {\bibfnamefont {U.}~\bibnamefont {Sassi}}, \bibinfo {author} {\bibfnamefont
  {A.~K.}\ \bibnamefont {Ott}}, \bibinfo {author} {\bibfnamefont {D.~D.}\
  \bibnamefont {Fazio}}, \bibinfo {author} {\bibfnamefont {D.}~\bibnamefont
  {Yoon}}, \bibinfo {author} {\bibfnamefont {M.}~\bibnamefont {Amado}},
  \bibinfo {author} {\bibfnamefont {A.~C.}\ \bibnamefont {Ferrari}}, \bibinfo
  {author} {\bibfnamefont {J.}~\bibnamefont {Linder}}, \ and\ \bibinfo {author}
  {\bibfnamefont {J.~W.~A.}\ \bibnamefont {Robinson}},\ }\href {\doibase
  10.1038/ncomms14024} {\bibfield  {journal} {\bibinfo  {journal} {Nature
  Communications}\ }\textbf {\bibinfo {volume} {8}},\ \bibinfo {pages} {14024}
  (\bibinfo {year} {2017})}\BibitemShut {NoStop}%
\bibitem [{\citenamefont {Zhang}\ \emph
  {et~al.}(2009{\natexlab{a}})\citenamefont {Zhang}, \citenamefont {Liu},
  \citenamefont {Qi}, \citenamefont {Dai}, \citenamefont {Fang},\ and\
  \citenamefont {Zhang}}]{zhang2009topological}%
  \BibitemOpen
  \bibfield  {author} {\bibinfo {author} {\bibfnamefont {H.}~\bibnamefont
  {Zhang}}, \bibinfo {author} {\bibfnamefont {C.-X.}\ \bibnamefont {Liu}},
  \bibinfo {author} {\bibfnamefont {X.-L.}\ \bibnamefont {Qi}}, \bibinfo
  {author} {\bibfnamefont {X.}~\bibnamefont {Dai}}, \bibinfo {author}
  {\bibfnamefont {Z.}~\bibnamefont {Fang}}, \ and\ \bibinfo {author}
  {\bibfnamefont {S.-C.}\ \bibnamefont {Zhang}},\ }\href
  {http://www.nature.com/nphys/journal/v5/n6/abs/nphys1270.html} {\bibfield
  {journal} {\bibinfo  {journal} {Nature Physics}\ }\textbf {\bibinfo {volume}
  {5}},\ \bibinfo {pages} {438} (\bibinfo {year}
  {2009}{\natexlab{a}})}\BibitemShut {NoStop}%
\bibitem [{\citenamefont {Liu}\ \emph {et~al.}(2010)\citenamefont {Liu},
  \citenamefont {Qi}, \citenamefont {Zhang}, \citenamefont {Dai}, \citenamefont
  {Fang},\ and\ \citenamefont {Zhang}}]{PhysRevB.82.045122}%
  \BibitemOpen
  \bibfield  {author} {\bibinfo {author} {\bibfnamefont {C.-X.}\ \bibnamefont
  {Liu}}, \bibinfo {author} {\bibfnamefont {X.-L.}\ \bibnamefont {Qi}},
  \bibinfo {author} {\bibfnamefont {H.}~\bibnamefont {Zhang}}, \bibinfo
  {author} {\bibfnamefont {X.}~\bibnamefont {Dai}}, \bibinfo {author}
  {\bibfnamefont {Z.}~\bibnamefont {Fang}}, \ and\ \bibinfo {author}
  {\bibfnamefont {S.-C.}\ \bibnamefont {Zhang}},\ }\href {\doibase
  10.1103/PhysRevB.82.045122} {\bibfield  {journal} {\bibinfo  {journal} {Phys.
  Rev. B}\ }\textbf {\bibinfo {volume} {82}},\ \bibinfo {pages} {045122}
  (\bibinfo {year} {2010})}\BibitemShut {NoStop}%
\bibitem [{\citenamefont {Hsieh}\ and\ \citenamefont
  {Fu}(2012{\natexlab{a}})}]{PhysRevLett.108.107005}%
  \BibitemOpen
  \bibfield  {author} {\bibinfo {author} {\bibfnamefont {T.~H.}\ \bibnamefont
  {Hsieh}}\ and\ \bibinfo {author} {\bibfnamefont {L.}~\bibnamefont {Fu}},\
  }\href {\doibase 10.1103/PhysRevLett.108.107005} {\bibfield  {journal}
  {\bibinfo  {journal} {Phys. Rev. Lett.}\ }\textbf {\bibinfo {volume} {108}},\
  \bibinfo {pages} {107005} (\bibinfo {year} {2012}{\natexlab{a}})}\BibitemShut
  {NoStop}%
\bibitem [{\citenamefont {Qi}\ \emph {et~al.}(2009)\citenamefont {Qi},
  \citenamefont {Hughes}, \citenamefont {Raghu},\ and\ \citenamefont
  {Zhang}}]{PhysRevLett.102.187001}%
  \BibitemOpen
  \bibfield  {author} {\bibinfo {author} {\bibfnamefont {X.-L.}\ \bibnamefont
  {Qi}}, \bibinfo {author} {\bibfnamefont {T.~L.}\ \bibnamefont {Hughes}},
  \bibinfo {author} {\bibfnamefont {S.}~\bibnamefont {Raghu}}, \ and\ \bibinfo
  {author} {\bibfnamefont {S.-C.}\ \bibnamefont {Zhang}},\ }\href {\doibase
  10.1103/PhysRevLett.102.187001} {\bibfield  {journal} {\bibinfo  {journal}
  {Phys. Rev. Lett.}\ }\textbf {\bibinfo {volume} {102}},\ \bibinfo {pages}
  {187001} (\bibinfo {year} {2009})}\BibitemShut {NoStop}%
\bibitem [{\citenamefont {Stone}\ and\ \citenamefont
  {Roy}(2004)}]{PhysRevB.69.184511}%
  \BibitemOpen
  \bibfield  {author} {\bibinfo {author} {\bibfnamefont {M.}~\bibnamefont
  {Stone}}\ and\ \bibinfo {author} {\bibfnamefont {R.}~\bibnamefont {Roy}},\
  }\href {\doibase 10.1103/PhysRevB.69.184511} {\bibfield  {journal} {\bibinfo
  {journal} {Phys. Rev. B}\ }\textbf {\bibinfo {volume} {69}},\ \bibinfo
  {pages} {184511} (\bibinfo {year} {2004})}\BibitemShut {NoStop}%
\bibitem [{\citenamefont {Michaeli}\ and\ \citenamefont {Fu}(2012)}]{MF12}%
  \BibitemOpen
  \bibfield  {author} {\bibinfo {author} {\bibfnamefont {K.}~\bibnamefont
  {Michaeli}}\ and\ \bibinfo {author} {\bibfnamefont {L.}~\bibnamefont {Fu}},\
  }\href {\doibase 10.1103/PhysRevLett.109.187003} {\bibfield  {journal}
  {\bibinfo  {journal} {Phys. Rev. Lett.}\ }\textbf {\bibinfo {volume} {109}},\
  \bibinfo {pages} {187003} (\bibinfo {year} {2012})}\BibitemShut {NoStop}%
\bibitem [{\citenamefont {Hor}\ \emph {et~al.}(2010)\citenamefont {Hor},
  \citenamefont {Williams}, \citenamefont {Checkelsky}, \citenamefont
  {Roushan}, \citenamefont {Seo}, \citenamefont {Xu}, \citenamefont
  {Zandbergen}, \citenamefont {Yazdani}, \citenamefont {Ong},\ and\
  \citenamefont {Cava}}]{Hetal10}%
  \BibitemOpen
  \bibfield  {author} {\bibinfo {author} {\bibfnamefont {Y.~S.}\ \bibnamefont
  {Hor}}, \bibinfo {author} {\bibfnamefont {A.~J.}\ \bibnamefont {Williams}},
  \bibinfo {author} {\bibfnamefont {J.~G.}\ \bibnamefont {Checkelsky}},
  \bibinfo {author} {\bibfnamefont {P.}~\bibnamefont {Roushan}}, \bibinfo
  {author} {\bibfnamefont {J.}~\bibnamefont {Seo}}, \bibinfo {author}
  {\bibfnamefont {Q.}~\bibnamefont {Xu}}, \bibinfo {author} {\bibfnamefont
  {H.~W.}\ \bibnamefont {Zandbergen}}, \bibinfo {author} {\bibfnamefont
  {A.}~\bibnamefont {Yazdani}}, \bibinfo {author} {\bibfnamefont {N.~P.}\
  \bibnamefont {Ong}}, \ and\ \bibinfo {author} {\bibfnamefont {R.~J.}\
  \bibnamefont {Cava}},\ }\href {\doibase 10.1103/PhysRevLett.104.057001}
  {\bibfield  {journal} {\bibinfo  {journal} {Phys. Rev. Lett.}\ }\textbf
  {\bibinfo {volume} {104}},\ \bibinfo {pages} {057001} (\bibinfo {year}
  {2010})}\BibitemShut {NoStop}%
\bibitem [{\citenamefont {Clogston}(1962)}]{C62}%
  \BibitemOpen
  \bibfield  {author} {\bibinfo {author} {\bibfnamefont {A.~M.}\ \bibnamefont
  {Clogston}},\ }\href {\doibase 10.1103/PhysRevLett.9.266} {\bibfield
  {journal} {\bibinfo  {journal} {Phys. Rev. Lett.}\ }\textbf {\bibinfo
  {volume} {9}},\ \bibinfo {pages} {266} (\bibinfo {year} {1962})}\BibitemShut
  {NoStop}%
\bibitem [{\citenamefont {Chandrasekhar}(1962)}]{C62b}%
  \BibitemOpen
  \bibfield  {author} {\bibinfo {author} {\bibfnamefont {B.~S.}\ \bibnamefont
  {Chandrasekhar}},\ }\href@noop {} {\bibfield  {journal} {\bibinfo  {journal}
  {Appl. Phys. Lett.}\ }\textbf {\bibinfo {volume} {1}},\ \bibinfo {pages} {7}
  (\bibinfo {year} {1962})}\BibitemShut {NoStop}%
\bibitem [{\citenamefont {Wray}\ \emph {et~al.}(2010)\citenamefont {Wray},
  \citenamefont {Xu}, \citenamefont {Xia}, \citenamefont {Qian}, \citenamefont
  {Fedorov}, \citenamefont {Lin}, \citenamefont {Bansil}, \citenamefont {Hor},
  \citenamefont {Cava},\ and\ \citenamefont {Hasan}}]{Wetal10}%
  \BibitemOpen
  \bibfield  {author} {\bibinfo {author} {\bibfnamefont {L.~A.}\ \bibnamefont
  {Wray}}, \bibinfo {author} {\bibfnamefont {S.}~\bibnamefont {Xu}}, \bibinfo
  {author} {\bibfnamefont {Y.}~\bibnamefont {Xia}}, \bibinfo {author}
  {\bibfnamefont {D.}~\bibnamefont {Qian}}, \bibinfo {author} {\bibfnamefont
  {A.~V.}\ \bibnamefont {Fedorov}}, \bibinfo {author} {\bibfnamefont
  {H.}~\bibnamefont {Lin}}, \bibinfo {author} {\bibfnamefont {A.}~\bibnamefont
  {Bansil}}, \bibinfo {author} {\bibfnamefont {Y.~S.}\ \bibnamefont {Hor}},
  \bibinfo {author} {\bibfnamefont {R.~J.}\ \bibnamefont {Cava}}, \ and\
  \bibinfo {author} {\bibfnamefont {M.~Z.}\ \bibnamefont {Hasan}},\ }\href
  {\doibase 10.1038/nphys1762} {\bibfield  {journal} {\bibinfo  {journal}
  {Nature Phys.}\ }\textbf {\bibinfo {volume} {6}},\ \bibinfo {pages} {855}
  (\bibinfo {year} {2010})}\BibitemShut {NoStop}%
\bibitem [{\citenamefont {Kriener}\ \emph {et~al.}(2011)\citenamefont
  {Kriener}, \citenamefont {Segawa}, \citenamefont {Ren}, \citenamefont
  {Sasaki}, \citenamefont {Wada}, \citenamefont {Kuwabata},\ and\ \citenamefont
  {Ando}}]{PhysRevB.84.054513}%
  \BibitemOpen
  \bibfield  {author} {\bibinfo {author} {\bibfnamefont {M.}~\bibnamefont
  {Kriener}}, \bibinfo {author} {\bibfnamefont {K.}~\bibnamefont {Segawa}},
  \bibinfo {author} {\bibfnamefont {Z.}~\bibnamefont {Ren}}, \bibinfo {author}
  {\bibfnamefont {S.}~\bibnamefont {Sasaki}}, \bibinfo {author} {\bibfnamefont
  {S.}~\bibnamefont {Wada}}, \bibinfo {author} {\bibfnamefont {S.}~\bibnamefont
  {Kuwabata}}, \ and\ \bibinfo {author} {\bibfnamefont {Y.}~\bibnamefont
  {Ando}},\ }\href {\doibase 10.1103/PhysRevB.84.054513} {\bibfield  {journal}
  {\bibinfo  {journal} {Phys. Rev. B}\ }\textbf {\bibinfo {volume} {84}},\
  \bibinfo {pages} {054513} (\bibinfo {year} {2011})}\BibitemShut {NoStop}%
\bibitem [{\citenamefont {Zhang}\ \emph
  {et~al.}(2009{\natexlab{b}})\citenamefont {Zhang}, \citenamefont {Qin},
  \citenamefont {Teng}, \citenamefont {Guo}, \citenamefont {Guo}, \citenamefont
  {Dai}, \citenamefont {Fang},\ and\ \citenamefont
  {Wu}}]{doi:10.1063/1.3200237}%
  \BibitemOpen
  \bibfield  {author} {\bibinfo {author} {\bibfnamefont {G.}~\bibnamefont
  {Zhang}}, \bibinfo {author} {\bibfnamefont {H.}~\bibnamefont {Qin}}, \bibinfo
  {author} {\bibfnamefont {J.}~\bibnamefont {Teng}}, \bibinfo {author}
  {\bibfnamefont {J.}~\bibnamefont {Guo}}, \bibinfo {author} {\bibfnamefont
  {Q.}~\bibnamefont {Guo}}, \bibinfo {author} {\bibfnamefont {X.}~\bibnamefont
  {Dai}}, \bibinfo {author} {\bibfnamefont {Z.}~\bibnamefont {Fang}}, \ and\
  \bibinfo {author} {\bibfnamefont {K.}~\bibnamefont {Wu}},\ }\href {\doibase
  10.1063/1.3200237} {\bibfield  {journal} {\bibinfo  {journal} {Applied
  Physics Letters}\ }\textbf {\bibinfo {volume} {95}},\ \bibinfo {pages}
  {053114} (\bibinfo {year} {2009}{\natexlab{b}})},\ \Eprint
  {http://arxiv.org/abs/http://dx.doi.org/10.1063/1.3200237}
  {http://dx.doi.org/10.1063/1.3200237} \BibitemShut {NoStop}%
\bibitem [{\citenamefont {Sandilands}\ \emph {et~al.}(2014)\citenamefont
  {Sandilands}, \citenamefont {Reijnders}, \citenamefont {Kriener},
  \citenamefont {Segawa}, \citenamefont {Sasaki}, \citenamefont {Ando},\ and\
  \citenamefont {Burch}}]{PhysRevB.90.094503}%
  \BibitemOpen
  \bibfield  {author} {\bibinfo {author} {\bibfnamefont {L.~J.}\ \bibnamefont
  {Sandilands}}, \bibinfo {author} {\bibfnamefont {A.~A.}\ \bibnamefont
  {Reijnders}}, \bibinfo {author} {\bibfnamefont {M.}~\bibnamefont {Kriener}},
  \bibinfo {author} {\bibfnamefont {K.}~\bibnamefont {Segawa}}, \bibinfo
  {author} {\bibfnamefont {S.}~\bibnamefont {Sasaki}}, \bibinfo {author}
  {\bibfnamefont {Y.}~\bibnamefont {Ando}}, \ and\ \bibinfo {author}
  {\bibfnamefont {K.~S.}\ \bibnamefont {Burch}},\ }\href {\doibase
  10.1103/PhysRevB.90.094503} {\bibfield  {journal} {\bibinfo  {journal} {Phys.
  Rev. B}\ }\textbf {\bibinfo {volume} {90}},\ \bibinfo {pages} {094503}
  (\bibinfo {year} {2014})}\BibitemShut {NoStop}%
\bibitem [{\citenamefont {Wray}\ \emph {et~al.}(2011)\citenamefont {Wray},
  \citenamefont {Xu}, \citenamefont {Xia}, \citenamefont {Qian}, \citenamefont
  {Fedorov}, \citenamefont {Lin}, \citenamefont {Bansil}, \citenamefont {Fu},
  \citenamefont {Hor}, \citenamefont {Cava},\ and\ \citenamefont
  {Hasan}}]{PhysRevB.83.224516}%
  \BibitemOpen
  \bibfield  {author} {\bibinfo {author} {\bibfnamefont {L.~A.}\ \bibnamefont
  {Wray}}, \bibinfo {author} {\bibfnamefont {S.}~\bibnamefont {Xu}}, \bibinfo
  {author} {\bibfnamefont {Y.}~\bibnamefont {Xia}}, \bibinfo {author}
  {\bibfnamefont {D.}~\bibnamefont {Qian}}, \bibinfo {author} {\bibfnamefont
  {A.~V.}\ \bibnamefont {Fedorov}}, \bibinfo {author} {\bibfnamefont
  {H.}~\bibnamefont {Lin}}, \bibinfo {author} {\bibfnamefont {A.}~\bibnamefont
  {Bansil}}, \bibinfo {author} {\bibfnamefont {L.}~\bibnamefont {Fu}}, \bibinfo
  {author} {\bibfnamefont {Y.~S.}\ \bibnamefont {Hor}}, \bibinfo {author}
  {\bibfnamefont {R.~J.}\ \bibnamefont {Cava}}, \ and\ \bibinfo {author}
  {\bibfnamefont {M.~Z.}\ \bibnamefont {Hasan}},\ }\href {\doibase
  10.1103/PhysRevB.83.224516} {\bibfield  {journal} {\bibinfo  {journal} {Phys.
  Rev. B}\ }\textbf {\bibinfo {volume} {83}},\ \bibinfo {pages} {224516}
  (\bibinfo {year} {2011})}\BibitemShut {NoStop}%
\bibitem [{\citenamefont {Li}\ \emph {et~al.}(2016)\citenamefont {Li},
  \citenamefont {Neupert}, \citenamefont {Bernevig},\ and\ \citenamefont
  {Yazdani}}]{JNBY16}%
  \BibitemOpen
  \bibfield  {author} {\bibinfo {author} {\bibfnamefont {J.}~\bibnamefont
  {Li}}, \bibinfo {author} {\bibfnamefont {T.}~\bibnamefont {Neupert}},
  \bibinfo {author} {\bibfnamefont {B.~A.}\ \bibnamefont {Bernevig}}, \ and\
  \bibinfo {author} {\bibfnamefont {A.}~\bibnamefont {Yazdani}},\ }\href@noop
  {} {\bibfield  {journal} {\bibinfo  {journal} {Nature Commun.}\ }\textbf
  {\bibinfo {volume} {7}},\ \bibinfo {pages} {10395} (\bibinfo {year}
  {2016})}\BibitemShut {NoStop}%
\bibitem [{\citenamefont {Nadj-Perge}\ \emph {et~al.}(2013)\citenamefont
  {Nadj-Perge}, \citenamefont {Drozdov}, \citenamefont {Bernevig},\ and\
  \citenamefont {Yazdani}}]{NDBY13}%
  \BibitemOpen
  \bibfield  {author} {\bibinfo {author} {\bibfnamefont {S.}~\bibnamefont
  {Nadj-Perge}}, \bibinfo {author} {\bibfnamefont {I.~K.}\ \bibnamefont
  {Drozdov}}, \bibinfo {author} {\bibfnamefont {B.~A.}\ \bibnamefont
  {Bernevig}}, \ and\ \bibinfo {author} {\bibfnamefont {A.}~\bibnamefont
  {Yazdani}},\ }\href {\doibase 10.1103/PhysRevB.88.020407} {\bibfield
  {journal} {\bibinfo  {journal} {Phys. Rev. B}\ }\textbf {\bibinfo {volume}
  {88}},\ \bibinfo {pages} {020407} (\bibinfo {year} {2013})}\BibitemShut
  {NoStop}%
\bibitem [{\citenamefont {Nadj-Perge}\ \emph {et~al.}(2014)\citenamefont
  {Nadj-Perge}, \citenamefont {Drozdov}, \citenamefont {Li}, \citenamefont
  {Chen}, \citenamefont {Jeon}, \citenamefont {Seo}, \citenamefont {MacDonald},
  \citenamefont {Bernevig},\ and\ \citenamefont {Yazdani}}]{Netal14}%
  \BibitemOpen
  \bibfield  {author} {\bibinfo {author} {\bibfnamefont {S.}~\bibnamefont
  {Nadj-Perge}}, \bibinfo {author} {\bibfnamefont {I.~K.}\ \bibnamefont
  {Drozdov}}, \bibinfo {author} {\bibfnamefont {J.}~\bibnamefont {Li}},
  \bibinfo {author} {\bibfnamefont {H.}~\bibnamefont {Chen}}, \bibinfo {author}
  {\bibfnamefont {S.}~\bibnamefont {Jeon}}, \bibinfo {author} {\bibfnamefont
  {J.}~\bibnamefont {Seo}}, \bibinfo {author} {\bibfnamefont {A.~H.}\
  \bibnamefont {MacDonald}}, \bibinfo {author} {\bibfnamefont {B.~A.}\
  \bibnamefont {Bernevig}}, \ and\ \bibinfo {author} {\bibfnamefont
  {A.}~\bibnamefont {Yazdani}},\ }\href@noop {} {\bibfield  {journal} {\bibinfo
   {journal} {Science}\ }\textbf {\bibinfo {volume} {346}},\ \bibinfo {pages}
  {602} (\bibinfo {year} {2014})}\BibitemShut {NoStop}%
\bibitem [{\citenamefont {Bravyi}\ and\ \citenamefont {Kitaev}(1988)}]{BK01}%
  \BibitemOpen
  \bibfield  {author} {\bibinfo {author} {\bibfnamefont {S.~B.}\ \bibnamefont
  {Bravyi}}\ and\ \bibinfo {author} {\bibfnamefont {A.~Y.}\ \bibnamefont
  {Kitaev}},\ }\href@noop {} {\enquote {\bibinfo {title} {Quantum codes on a
  lattice with boundary},}\ } (\bibinfo {year} {1988}),\ \bibinfo {note}
  {preprint},\ \Eprint {http://arxiv.org/abs/quant-ph/9811052}
  {arXiv:quant-ph/9811052} \BibitemShut {NoStop}%
\bibitem [{\citenamefont {Vijay}\ \emph {et~al.}(2015)\citenamefont {Vijay},
  \citenamefont {Hsieh},\ and\ \citenamefont {Fu}}]{VHF15}%
  \BibitemOpen
  \bibfield  {author} {\bibinfo {author} {\bibfnamefont {S.}~\bibnamefont
  {Vijay}}, \bibinfo {author} {\bibfnamefont {T.~H.}\ \bibnamefont {Hsieh}}, \
  and\ \bibinfo {author} {\bibfnamefont {L.}~\bibnamefont {Fu}},\ }\href
  {\doibase 10.1103/PhysRevX.5.041038} {\bibfield  {journal} {\bibinfo
  {journal} {Phys. Rev. X}\ }\textbf {\bibinfo {volume} {5}},\ \bibinfo {pages}
  {041038} (\bibinfo {year} {2015})}\BibitemShut {NoStop}%
\bibitem [{\citenamefont {Vijay}\ and\ \citenamefont {Fu}(2016)}]{VF15}%
  \BibitemOpen
  \bibfield  {author} {\bibinfo {author} {\bibfnamefont {S.}~\bibnamefont
  {Vijay}}\ and\ \bibinfo {author} {\bibfnamefont {L.}~\bibnamefont {Fu}},\
  }\href {\doibase 10.1088/0031-8949/T168/1/014002} {\bibfield  {journal}
  {\bibinfo  {journal} {Phys. Scr.}\ }\textbf {\bibinfo {volume} {T168}},\
  \bibinfo {pages} {014002} (\bibinfo {year} {2016})}\BibitemShut {NoStop}%
\bibitem [{\citenamefont {Hsieh}\ and\ \citenamefont
  {Fu}(2012{\natexlab{b}})}]{HF12}%
  \BibitemOpen
  \bibfield  {author} {\bibinfo {author} {\bibfnamefont {T.~H.}\ \bibnamefont
  {Hsieh}}\ and\ \bibinfo {author} {\bibfnamefont {L.}~\bibnamefont {Fu}},\
  }\href {\doibase 10.1103/PhysRevLett.108.107005} {\bibfield  {journal}
  {\bibinfo  {journal} {Phys. Rev. Lett.}\ }\textbf {\bibinfo {volume} {108}},\
  \bibinfo {pages} {107005} (\bibinfo {year} {2012}{\natexlab{b}})}\BibitemShut
  {NoStop}%
\end{thebibliography}%
\pagebreak
\appendix

\onecolumngrid

\setcounter{equation}{0}
\setcounter{figure}{0}

\renewcommand{\theequation}{A\arabic{equation}}
\renewcommand{\thefigure}{A\arabic{figure}}

\renewcommand{\bibnumfmt}[1]{[#1]}
\renewcommand{\citenumfont}[1]{#1}

\section{Derivation of the Projected Hamiltonian}
In this section, we derive the projected Hamiltonian describing the two-dimensional $p$-wave topological superconductor used in the main text. A simple four-band model for the valence and conduction bands of the topological insulator Bi$_2$Se$_3$ near $\Gamma$ is \cite{zhang2009topological, PhysRevB.82.045122, HF12}
\begin{align}
{\cal H}^\text{3D}(\mathbf{k}) &= E_G \sigma_x + v_z k_z \sigma_y + v \sigma_z \left( k_x s_y - k_y s_x \right),
\label{hamil3D}
\end{align}
where parameters $E_G,$ $v_z, $ and $v$ are determined from experimental data. The four bands correspond to the two dominant Wannier orbitals  $\{ A , B \}$, each with spin $ \{\uparrow , \downarrow \}$. The $2 \times 2$ Pauli matrices $\sigma_i$ and $s_i$, with $i = x, y, z,$ act on orbital and spin spaces respectively. We use units where $\hbar = 1$, so that all dimensions can be expressed in terms of energies and lengths.

The bands are doubly degenerate with dispersion
\begin{align}
E(\mathbf{ k}) & =\pm  \sqrt{E_G^2 + v_z^2 k_z^2 + v^2 ( k_x^2 + k_y^2 )}\ .
\end{align}
The gap between the conduction band and the valence band is thus $2 E_G$. 
In order to describe a superconductor, we need to extend the Hilbert space to include hole wavefunctions and a corresponding hole Hamiltonian. Using the convention given in Ref.~\cite{HF12}, the full wavefunction now contains eight components,
\begin{align}
\left| \Psi \right\rangle &= \left( \begin{array}{cccccccccccc} \psi_{ A, \uparrow}^{(e)} & \psi_{ A, \downarrow}^{(e)}  & \psi_{ B, \uparrow}^{(e)} & \psi_{ B, \downarrow}^{(e)} & - \psi_{ A, \downarrow}^{(h)} & \psi_{ A, \uparrow}^{(h)} & - \psi_{ B, \downarrow}^{(h)} & \psi_{B, \uparrow}^{(h)} \end{array} \right)^T.
\end{align}
In this basis, the $8\times8$ Hamiltonian is simply 
\begin{equation}
\mathcal{H}(\mathbf{{k}})=(  {\cal H}^\text{3D}(\mathbf{{k}}) - {\epsilon}_F ) \tau_z,
\end{equation}
 where $\epsilon_F$ is the Fermi energy, and $\boldsymbol{\tau}$ is a third set of Pauli matrices that
couples particles and holes of the same spin and in the same orbital. The particles and holes are now mixed by a triplet, time-reversal-invariant coupling  of the form $\Delta = \Delta_\text{sc} \sigma_y s_z \tau_x$ that induces
a superconducting gap proportional to $\Delta_\text{sc}$.

If the energy gap is large, i.e., $2E_G \gtrsim \epsilon_F - E_G$ and $E_G \gg \Delta_\text{sc}$, only the lower part of the conduction band is occupied and we can neglect the valence band, as the conduction and valence bands are separated by a large energy gap $2 E_G$. To simplify the result, we first perform a rotation in orbital space by $\pi / 2 $ about the $y$ axis, so that $\sigma_y \rightarrow \sigma_y , \sigma_x \rightarrow \sigma_z , \sigma_z \rightarrow - \sigma_x$. In this new representation, the states at the edge of the conduction and valence bands are eigenstates of $\sigma_z$, with eigenvalues $\pm 1$. The operators ${\cal P}_{\pm} = ( {\cal I} \pm \sigma_z ) / 2$, where ${\cal I}$ is the identity matrix, thus project the wavefunctions onto the conduction and valence bands respectively. The kinetic energy in Eq.~(\ref{hamil3D}), which is now proportional to $\sigma_x$, and the gap, which remains proportional to $\sigma_y$, mix the conduction and valence bands. Using perturbation theory, we can expand the Hamiltonian to first order in $\Delta_\text{sc}$ and then project onto the conduction band to find the reduced Hamiltonian
\begin{align}
{\cal H}^\text{red}(\mathbf{ k}) &\approx  {\cal P}_+ \left\{ \left[ v_F \left( k_x s_y - k_y s_x \right) \sigma_x \tau_z + v_z k_z \sigma_y \right] \left( E_G \sigma_z \tau_z \right)^{-1} v_F \left( k_x s_y - k_y s_x \right) \sigma_x \tau_z \right\} \mathcal{P}_+  \nonumber \\ &-  {\cal P}_+ \left\{ v_F \left[ \left( k_x s_y - k_y s_x \right) \sigma_x \tau_z + v_z k_z \sigma_y \right] \left( E_G \sigma_z \tau_z \right)^{-1} \Delta_\text{sc} \sigma_y s_z \tau_x 
\right\} \mathcal{P}_+ \nonumber \\ &- {\cal P}_+ \left\{ \Delta_\text{sc} \sigma_y s_z \tau_x \left( E_G \sigma_z \tau_z \right)^{-1} \left[ v_F \left( k_x s_y - k_y s_x \right) \sigma_x \tau_z + v_z k_z \sigma_y \right]\right\}\mathcal{P}_+. 
\label{hamilproy}
\end{align}
Using $\sigma_z^{-1} = \sigma_z , \tau_z^{-1} = \tau_z$, and adding the chemical potential that was omitted in Eq.~(\ref{hamilproy}), we obtain
\begin{equation}
{\cal H}^{\text{red}}(\mathbf{ k}) = \left[ \left( \frac{v_F^2 | \mathbf{ k} |^2 + v_z^2 k_z^2}{2 E_G} - \bar{\epsilon}_F \right) \tau_z  - \frac{2 v_F \Delta_\text{sc}}{E_G} \tau_x \left( k_x s_x + k_y s_y \right)  - \frac{2 v_z \Delta_\text{sc}}{E_G} \tau_x k_z s_z \right] \mathcal{P}_+    ,
\label{hamilf}
\end{equation}
where $\bar{\epsilon}_F = \epsilon_F - E_G$.

This Hamiltonian defines a three-dimensional topological superconductor. The Fermi 
wave-vector and the coherence length along to the $z$-direction are given by
\begin{align}
k_{F z} &= \sqrt{\frac{2 E_G \bar{\epsilon}_F}{v_z^2}}  \nonumber ,\\
\xi_z^{-1} &= \frac{2 \Delta_\text{sc}}{v_z}.
\end{align}
We now analyze a quasi-two-dimensional system of thickness $d$, where $k_{Fz}^{-1}  \ll d \lesssim \xi_z$. In the absence of superconductivity, the Hamiltonian can be separated as ${\cal H}^\text{red} = {\cal H}_\parallel + {\cal H}_z$, where ${\cal H}_\parallel$ and ${\cal H}_z$ are the in-plane and out-of-plane components of the Hamiltonian, respectively. The electronic states can then be divided into sub-bands, $\epsilon_n ( {\bf k}) = \epsilon_z^n + \epsilon_\parallel ( {\bf k} )$. Here, $\epsilon_{z}^n$ and $\epsilon_\parallel ( {\bf k} )$ are eigenvalues of ${\cal H}_z$ and ${\cal H}_\parallel$, respectively, and $\mathbf{k}$ is the two-dimensional wave-vector in the $x$-$y$ plane. The separation of levels due to the quantization in the $z$-direction is, approximately, 
\begin{align}
\delta \epsilon_z^n &\approx \frac{v_z^2 k_{F z}}{E_G d} \gg \frac{v_z k_{F z} \Delta_\text{sc}}{E_G} \approx \Delta_\text{exp},
\end{align}
where $\Delta_\text{exp}$ is the value of the measured superconducting gap at the Fermi energy, see Eq.~(\ref{hamilf}). In order to define a low-energy effective theory for the electronic structure, we study first the lowest energy state for $|  {\bf k} | \approx 0$ and energy $| \epsilon |  \sim \Delta_\text{sc} \ll \delta \epsilon_z^n$.

 For $| {\bf k} | = 0$, the Hamiltonian in Eq.~(\ref{hamilf}) can be split into two $2 \times 2$ Hamiltonians, one which mixes electrons and holes with spin up, and another one which mixes electrons and holes with spin down. The two Hamiltonians are related by time-reversal symmetry. For $d \gtrsim \xi_z$ the lowest states of these Hamiltonians are localized at the top and bottom surfaces. The decay length of these states is given by the coherence length $\xi_z$. 
 When $d \lesssim \xi_z$, these states hybridize, and give rise to two Andreev states at energies $\pm \epsilon_0 \lesssim \Delta_\text{sc}$. Thus, the combination of the two $2 \times 2$ Hamiltonians gives rise to four low-energy states, $| i \rangle , i = 1 , \cdots , 4$, related by particle-hole and time-reversal symmetries.

 A generic wavefunction in this low-energy sector can be written as $\sum_i \psi_i ( \mathbf{r} ) | i \rangle$, where $\mathbf{r}$ is a two dimensional vector. Using the basis of states $| i \rangle$, we can write an effective Hamiltonian whose states are four-component spinors. Without loss of generality, we can  label the two states at energy $+ \epsilon_0$ as $| e \uparrow \rangle$ and $| e \downarrow \rangle$, and the two spinors with energy $- \epsilon_0$ as $| h \uparrow \rangle$ and $| h \downarrow \rangle$. 
 We assume that the matrix elements of the spin Pauli matrices $s_{i }$  and the electron-hole Pauli matrices $\tau_{i}$ ($i=x,y,z$) have the same values in this basis as in the original basis of spins and electron-hole states. Hence, we can write the effective Hamiltonian for the in-plane wavefunction in terms of $x$ and $y$ coordinates as
\begin{align}
{\cal H}^\text{2D} (\mathbf{ k}) &=   \left[ \left( \frac{v_F^2 | \mathbf{ k} |^2 }{2 E_G} - \bar{\epsilon}_F   - \epsilon_0  \right) \tau_z  - \frac{2 v_F \Delta_\text{sc}{\rm sign} ( \epsilon_0 )}{E_G} \tau_x \left( k_x s_x + k_y s_y \right) \right].
\label{hamilf2D}
\end{align}
This Hamiltonian can be split into two independent Hamiltonians related by time-reversal symmetry. These Hamiltonians mix electrons with spin up and holes with spin down, and vice versa. 

We make an approximate decomposition into a state
with small $|\mathbf{k}|$, and a lowest energy $k_z$ state.
To good approximation, the Hamiltonian depends only on $k_z$, 
\begin{equation}
\mathcal{H}_z = \bar{\epsilon}_F\left( \begin{array}{ccc}
\frac{k_z^2}{k_{Fz}^2}-1 & -\frac{2k_z}{\xi_z k_{Fz}^2}  \\
-\frac{2k_z}{\xi_z k_{Fz}^2}  & -\frac{k_z^2}{k_{Fz}^2}+1  \end{array} \right).
\end{equation}
A standard plane-wave solution to this Hamiltonian has eigenvalues 
\begin{equation}
\left(\frac{\epsilon_0}{\bar{\epsilon}_F}\right)^2=
\left( \frac{k_z^2}{k_{Fz}^2} - 1 \right)^2   
+\left(  \frac{2 k_z}{\xi_z k_{Fz}^2}\right)^2  . \label{eq:eval}
\end{equation}
This spectrum has an energy gap of about $4/\xi_zk_{Fz}$ due to superconductivity.

\begin{figure}
\begin{center}
\includegraphics[width=8cm]{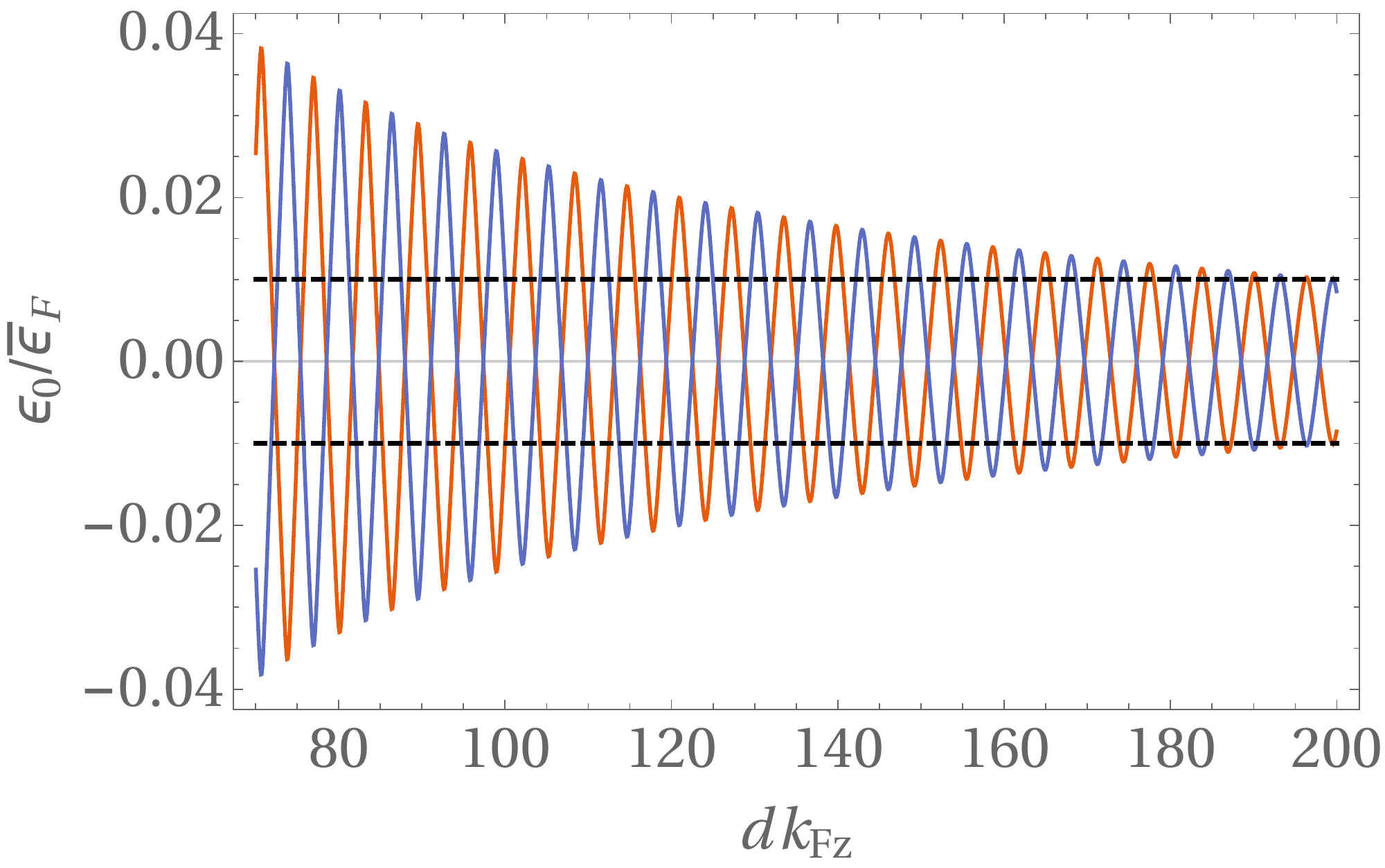}
\end{center}
\caption{Plots of the  energy  $\epsilon_0$ in Eq.~(\ref{hamilf2D}) as function of the thickness of the disc $d$. The dashed line shows the gap $\pm 2/(k_{Fz}\xi_z)$. We have used $\xi_z=4000~\mathrm{\AA}$ and  $k_{F z}^{-1} = 20~\mathrm{\AA}$. This choice of parameters keeps the superconducting gap fixed at 0.6 meV.
}
\label{fig:width_disk}
\end{figure}

We can solve for the eigenenergies; 
values of the parameter $\epsilon_0$ in Eq.~(\ref{hamilf2D})  are shown in Fig.~\ref{fig:width_disk} for a realistic choice of parameters. Here we require that the superconducting gap is 0.6 meV. For energy eigenvalues inside the gap, $\epsilon_0/\bar{\epsilon}_F < 2/\xi_zk_{Fz},$ we have states that are weakly localized at the top and bottom surfaces, with an exponential decay slower than the thickness of the disc. As illustrated in Fig.~\ref{fig:width_disk}, there is a wide range of values for $d$ for which we find a state inside or near the gap.

 We note that for  $k_{F z}^{-1} = 20~\mathrm{\AA}$, we have $\xi_z=4000~\mathrm{\AA}$ (a smaller Fermi wavevector leads to an even larger coherence length for a fixed gap). This shows that our system needs not be atomistically thin, but can have a thickness on the order of a  thousand angstroms,  similar to the coherence length. Henceforth, without loss of generality, we assume we have chosen a particular thickness $d$ for the disc where no localization at the surfaces occurs. We also absorb $\epsilon_0$ into the definition of $\bar{\epsilon}_F$. This does not change the value of $\bar{\epsilon}_F$ appreciably because $\bar{\epsilon}_F$ is much larger than $\epsilon_0.$

The two reduced Hamiltonians 
can be interpreted as describing two metals with quadratic dispersion and superconducting gaps at the Fermi surface with angular dependence $\Delta_\text{sc}^\text{Fermi} =2  v_F k_F \Delta_\text{sc} e^{i \varphi} / E_G    $, where $\varphi = \arctan ( k_y / k_x )$ is the angular coordinate in momentum space, and $k_F = \sqrt{ 2 E_G  \bar{\epsilon}_F  / v_F^2}$. In experiments, the more relevant energy scale is the superconducting energy gap measured at the Fermi  surface $2\Delta_\text{exp} = 2|\Delta_\text{sc}^\text{Fermi}| = 4v_Fk_F\Delta_\text{sc}/E_G .$ The two Hamiltonians are
\begin{align}
{\cal H}^{\text{red}}_{\pm} ( \mathbf{ k} ) &= \left( \begin{array}{cc} \frac{v_F^2 ( k_x^2 +  k_y^2 )}{2 E_G} - \bar{\epsilon}_F & \frac{2 v_F \Delta_\text{sc} (-i  k_x \pm k_y )}{E_G} \\ \frac{2 v_F \Delta_\text{sc} ( i k_x \pm k_y )}{E_G} &- \frac{v_F^2 ( k_x^2 + k_y^2 )}{2 E_G} + \bar{\epsilon}_F \end{array} \right).
\label{hamilf2}
\end{align}

The two Hamiltonians are related by time inversion. Each Hamiltonian describes a chiral $p$-wave superconductor \cite{RG00}. The energy dispersion is
\begin{align}
E(\mathbf{ k}) &= \pm \sqrt{\left( \frac{v_F^2 ( k_x^2 +  k_y^2 )}{2 E_G} - \bar{\epsilon}_F \right)^2 + \left( \frac{2 v_F \Delta_\text{sc}}{E_G} \right)^2 \left( k_x^2 + k_y^2 \right)}.
\end{align}
Using the convention  where 
\begin{equation}
|\Psi\rangle = \left( \begin{array}{cccc}
\psi^{(e)}_\uparrow & \psi^{(h)}_\downarrow  & -\psi^{(h)}_\uparrow &\psi^{(e)}_\downarrow \\  \end{array} \right)^T,
\label{eq:psiorder}
\end{equation}
 the full Hamiltonian should be written as 
\begin{equation}
\label{eq: fullhamiltonian}
\mathcal{H}_0(\mathbf{k}) = \left( \begin{array}{ccc}
\mathcal{H}_+(\mathbf{k}) & 0  \\
0 & \mathcal{H}_-(\mathbf{k})   \end{array} \right),
\end{equation}
where
\begin{align}
{\cal H}_{+} ( \mathbf{ k} ) &= \left( \begin{array}{cc}  \frac{v_F^2 ( k_x^2 +  k_y^2 )}{2 E_G} - \bar{\epsilon}_F & \frac{2 v_F \Delta_\text{sc} (-i  k_x + k_y )}{E_G} \\  \frac{2 v_F \Delta_\text{sc} ( i k_x + k_y )}{E_G} &- \frac{v_F^2 ( k_x^2 + k_y^2 )}{2 E_G} + \bar{\epsilon}_F \end{array} \right),
\label{hamilf3}
\end{align}
and ${\cal H}_{-}( \mathbf{ k} )=- {\cal H}_{+}(- \mathbf{ k} )$.

\section{Edge States on a Disc}

In this section, we solve for the edge modes on a disc. As noted above, the low-energy states in the conduction band of the two-dimensional superconductor doped Bi$_2$Se$_3$ can be described by two effective $p$-wave Hamiltonians that couple electrons and holes of opposite spins defined by Eqs.~(\ref{eq: fullhamiltonian}) and~(\ref{hamilf3}). The $p^+$ and $p^-$ superconducting branches are related by time-reversal symmetry. Thus, since eigenstates of $\mathcal{H}_+$ are also eigenstates of $\mathcal{H}_-$, we restrict our attention to $\mathcal{H}_+$ on a circle of radius $R$. In polar coordinates, we have $\pm ik_x+k_y = \pm e^{\pm i\theta} \left( \partial_r \mp i\partial_\theta /r \right)$, and $k_x^2+k_y^2 = - \nabla^2 = - \partial_r^2-\partial_r/r-\partial^2_\theta/r^2$. Using separation of variables, we can write our wavefunction in coordinate space as
\begin{equation}
\label{eq: simplified psi}
\Psi_\ell(r, \theta) = e^{i\ell\theta}\left( \begin{array}{ccc}
e^{i\theta}\psi_{+,\ell}(r)  \\
\psi_{-,\ell}(r) \end{array}  \right).
\end{equation}
To insist that the wavefunction is single-valued, we have that $\ell$ must be an integer. In this case, $\ell$ takes the role of describing our angular momentum quantum number. Let us now show that indeed, $\ell$ can be used as a good quantum number. The angular momentum operator in this case is
\begin{equation}
L_z = -i\partial_\theta \mathcal{I} - \frac{1}{2} s_z = \left( \begin{array}{ccc}
\frac{\partial_\theta}{i}-\frac{1}{2} & 0  \\
0 & \frac{\partial_\theta}{i}+\frac{1}{2}  \end{array} \right),
\label{angular}
\end{equation}
where $\mathcal{I}$ is the identity matrix. The spectrum of $L_z$ is $L_z\Psi_\ell = \left(\ell+\frac{1}{2} \right)\Psi_\ell$. By using the fact that partial derivatives commute, it immediately follows that $[\mathcal{H}_+, L_z] = 0$. Consequently, we can diagonalize both $\mathcal{H}_+$ and $L_z$ simultaneously, and use $\ell$ as a good quantum number. 

Using the form of $\Psi$ in Eq.~(\ref{eq: simplified psi}), we find that the radial wavefunctions satisfy the following eigenvalue problem
\begin{equation}
\label{eq: eigenvalue problem for radial components}
\left( \begin{array}{ccc}
 -\frac{v_F^2}{2 E_G} \left(\partial_r^2+\frac{\partial_r}{r} -\frac{(\ell+1)^2}{r^2} \right)-\bar{\epsilon}_F & -\frac{2v_F \Delta_\text{sc}}{E_G} \left(\partial_r-\frac{\ell}{r} \right)  \\
\frac{2v_F \Delta_\text{sc}}{E_G} \left( \partial_r + \frac{(\ell+1)}{r} \right) & \frac{v_F^2}{2 E_G} \left(\partial_r^2+\frac{\partial_r}{r} -\frac{\ell^2}{r^2} \right)+\bar{\epsilon}_F   \end{array} \right)\left( \begin{array}{ccc}
\psi_{+,\ell}(r) \\
\psi_{-,\ell}(r)  \end{array} \right) = E\left( \begin{array}{ccc}
\psi_{+,\ell}(r) \\
\psi_{-,\ell}(r)  \end{array} \right).
\end{equation}
To solve for the radial wavefunction, we make an ansatz that the wavefunction components are Bessel functions based on the motivation that eigenfunctions of the Laplacian are Bessel functions
\begin{equation}
\psi_{+,\ell}(r) = aJ_{\ell+1}(kr), \quad \text{and} \quad \psi_{-,\ell}(r) = bJ_{\ell}(kr),
\end{equation}
where $a$ and $b$ are constants, $J_\ell(x)$ is the Bessel function of the first kind of order $\ell,$ and $k$ is the linear momentum in the radial direction. Substituting this ansatz into Eq.~(\ref{eq: eigenvalue problem for radial components}), our problem is reduced to a simple eigenvalue problem 
\begin{equation}
\left( \begin{array}{ccc}
\frac{v_F^2}{2 E_G} k^2-\bar{\epsilon}_F & \frac{2v_F \Delta_\text{sc}}{E_G}  k \\
\frac{2v_F \Delta_\text{sc}}{E_G} k & - \frac{v_F^2}{2 E_G} k^2 +\bar{\epsilon}_F \end{array} \right)\left( \begin{array}{ccc}
a \\
b\end{array} \right) = E\left( \begin{array}{ccc}
a \\
b\end{array} \right).
\end{equation}
This has unnormalized solutions 
\begin{equation}
\Psi_\ell(r,\theta) =  e^{i\ell \theta}\left( \begin{array}{ccc}
-\frac{2v_F \Delta_\text{sc}}{E_G} k e^{i\theta} J_{\ell+1}(kr)  \\
\left( \frac{v_F^2}{2 E_G} k^2-\bar{\epsilon}_F -E \right)J_\ell(kr)  \end{array} \right) ,
\end{equation}
where 
\begin{equation}
E = \pm \sqrt{\left( \frac{v_F^2}{2 E_G} k^2-\bar{\epsilon}_F  \right)^2+\left( \frac{2v_F \Delta_\text{sc}}{E_G} k \right)^2}.
\end{equation}
Having found the eigenstates of our topological superconductor, we can now study the effect of the edge. We impose hard-wall boundary condition whereby we require the wavefunction to vanish at the perimeter of the disc $\Psi_\ell(R,\theta) = 0$. This condition is equivalent to continuity of the Dirac wavefunction at the boundary where we have Andreev reflections \cite{PhysRevB.69.184511}. In order to satisfy this condition, the wavefunction for a particular energy $E$ must be a linear combination of outgoing and reflected waves. 

Let us now simplify by using dimensionless variables, where we express lengths in terms of the
Fermi momentum $k_F=\sqrt{2E_G\bar{\epsilon}_F}/v_F$ as defined in the main text, and all energies in terms of $\bar{\epsilon}_F$,
\begin{equation}
\label{eq: kappa}
\kappa_\pm^2 = \left(  \frac{k_\pm}{k_F} \right)^2 = 1-\gamma \pm \sqrt{\gamma^2-2\gamma+\mathcal{E}^2},
\end{equation}
where $\gamma = \frac{4 \Delta_\text{sc} ^2}{E_G \bar{\epsilon }_F} = \frac{\Delta_\text{exp}^2}{2\bar{\epsilon}_F^2}\ll 1$ measures the superconducting gap relative to $E_G$ and $\bar{\epsilon}_F$ and $\mathcal{E} = E/\bar{\epsilon }_F$. In terms of these parameters, the full wavefunction is
\begin{equation}
\label{eq: wavefunction}
\begin{split}
\Psi_{\mathcal{E},\ell}(\rho,\theta) &= C_+e^{i\ell\theta}\left( \begin{array}{ccc}
-\sqrt{2\gamma} \kappa_+ e^{i\theta} J_{\ell+1}(\lambda\kappa_+  \rho) \\
\left( \kappa_+^2-1-\mathcal{E} \right)J_\ell(\lambda\kappa_+  \rho)\end{array} \right)-C_-e^{i\ell\theta}\left( \begin{array}{ccc}
-\sqrt{2\gamma}  \kappa_- e^{i\theta} J_{\ell+1}(\lambda\kappa_-  \rho) \\
\left( \kappa_-^2-1-\mathcal{E} \right)J_\ell(\lambda\kappa_-  \rho)\end{array} \right),
\end{split}
\end{equation}
where $\rho=r/R$ is the fractional radius, and $\lambda = k_F R$ is the dimensionless system size. Setting both components of the above equation to zero at $\rho=1$, we obtain the following transcendental equation for the allowed energies
\begin{equation}
\begin{split}
\label{eq: transcendental}
\kappa_+\left( \kappa_-^2-1-\mathcal{E} \right)J_\ell(\lambda \kappa_-)J_{\ell+1}(\lambda \kappa_+)-\kappa_-\left( \kappa_+^2-1-\mathcal{E} \right)J_\ell(\lambda \kappa_+)J_{\ell+1}(\lambda \kappa_-)=0.
\end{split}
\end{equation}
We can numerically solve this transcendental equation for some typical parameter values, given that we stay within the regime of validity for our effective Hamiltonian.  To obtain an edge state with energy below the gap, we can consider the limit of a very large disc, as done in \cite{PhysRevB.69.184511}. Suppose that $\lambda \gg 1$, then for a very small angle $\theta \ll 1$, the boundary resembles that of the infinite half plane for which we know that the edge state has energy $\mathcal{E}^\text{edge} = -\frac{2v_F k_x \Delta_\text{sc}}{E_G\bar{\epsilon}_F}$, where $k_x$ is the linear momentum along the edge \cite{PhysRevB.69.184511, FFMN07}. Translating this result to our geometry in the semi-classical limit, we have $|\mathbf{L}| = |\mathbf{r}\times \mathbf{p}| = \ell  \approx Rk_x. $
Thus, approximately, the energy of the edge state in the semi-classical limit is $\mathcal{E}_\text{edge} \approx -\frac{ \ell }{\ell_0}\delta$, where $\delta = \Delta_\text{sc}/\bar{\epsilon}_F$ and $ \ell_0 = R E_G/2v_F$. We can tighten this estimate further using a general argument of particle-hole symmetry; that is,  if $\mathcal{H}\Psi = E\Psi$, then $\mathcal{H}(\tau_x\mathcal{C}\Psi)=-E(\tau_x\mathcal{C}\Psi),$ where $\mathcal{C}$ is the complex-conjugation operator. With this, we see that a state of angular momentum $\ell$ has the opposite energy to a state of angular momentum $-(\ell+1)$. Therefore, we find that the edge energy is 
\begin{equation}
\label{eq: positive edge energy}
\mathcal{E}_{+}^\text{edge} = -\frac{ \ell +\frac{1}{2} }{\ell_0}\delta = - \left( \ell +\frac{1}{2}\right) \frac{\sqrt{2\gamma}}{\lambda}.
\end{equation}
This can be further improved by a more detailed asymptotic analysis. We find
\begin{equation}
\label{eq: positive edge energy2}
\mathcal{E}_{+}^\text{edge}=-\frac{\sqrt{\gamma/2 }}{\lambda} (2 \ell+1) \left(
1+
\frac{\sqrt{\gamma/2 }}{\lambda}
+\frac{\gamma -\frac{1}{2}}{\lambda^2}
\right)+\mathcal{O}(\lambda^{-4}),
\end{equation}
with a numerical indication that the next order has a more complex dependence on $\ell$.

\begin{figure}
\begin{center}
\includegraphics[width=8cm]{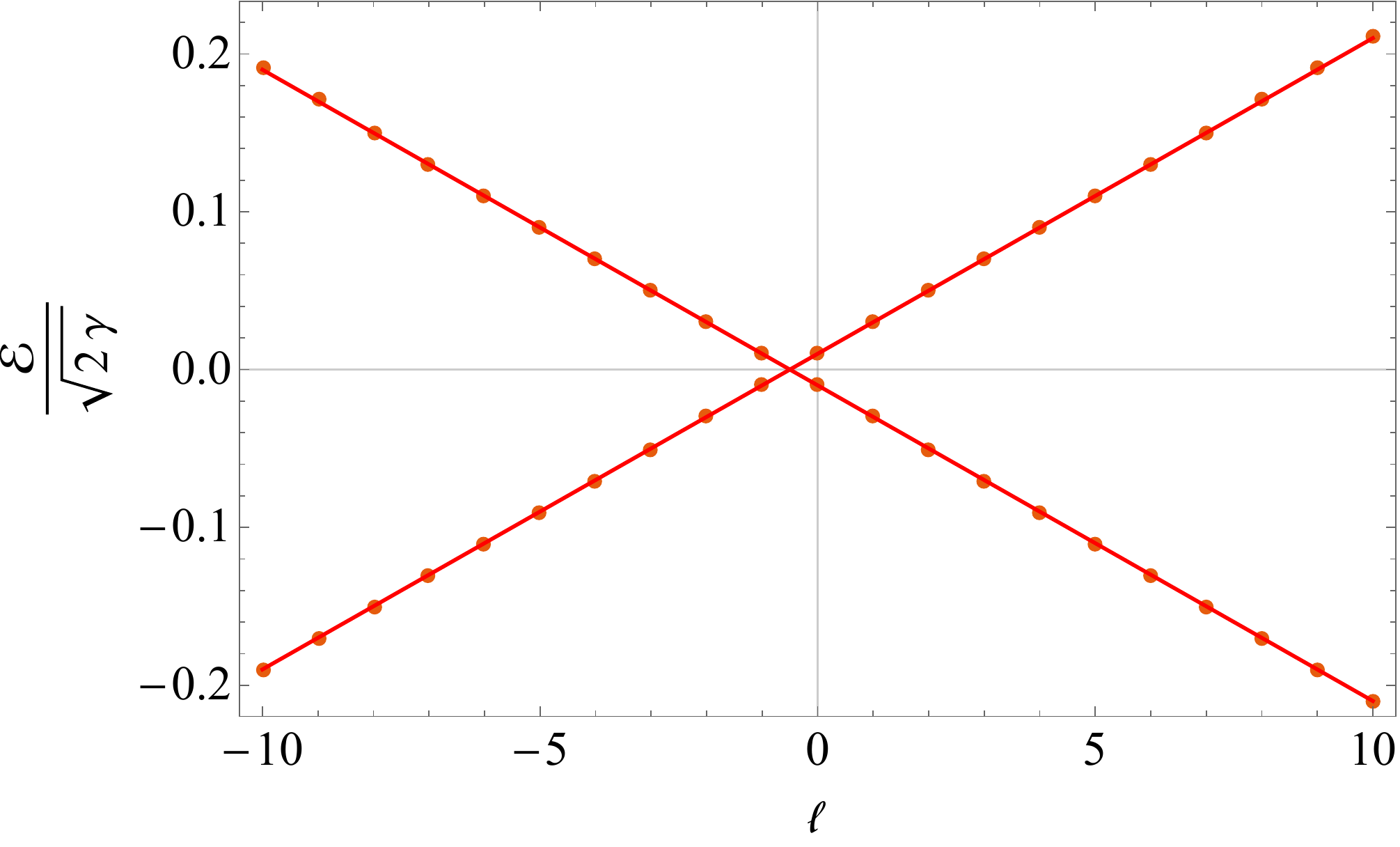}
\end{center}
\caption{Spectrum of the edge states for $\mathcal{H}_+$ and $\mathcal{H}_-$ with $\gamma = 1/16$ and $\lambda = 50$. The dots are obtained from numerical evaluation of Eq.~(\ref{eq: transcendental}), and the solid line is plotted using Eq.~(\ref{eq: positive edge energy}). The agreement between Eq.~(\ref{eq: positive edge energy}) and numerical computation is good beyond the domain shown here.}
\label{fig: spectrum of edge states appen}
\end{figure}

\begin{figure}
\begin{center}
\includegraphics[width= 8cm]{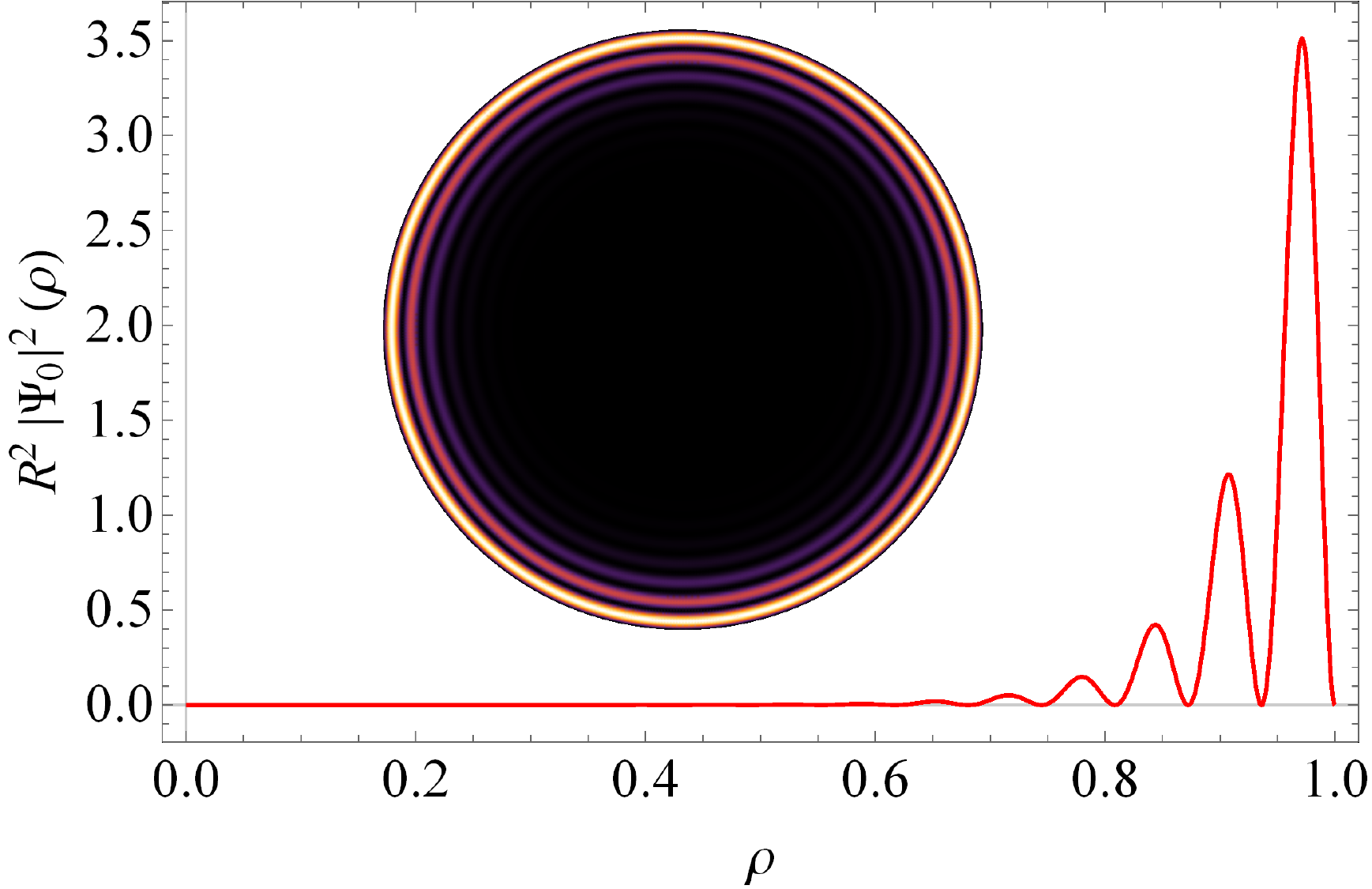}
\end{center}
\caption{Probability density of an edge mode for $\ell = 0$ as a function of $\rho$. The inset shows the same probability density distributed on the disc. The parameters used here are $\gamma = 1/16$ and $\lambda = 50$.}
\label{fig: localized mode appen}
\end{figure}

We compare in Fig.~\ref{fig: spectrum of edge states appen} the functional form of Eq.~(\ref{eq: positive edge energy}) to the lowest energy found from the transcendental equation in Eq.~(\ref{eq: transcendental}). To leading order in $\lambda^{-1}$, there is good agreement between the numerical result and the approximation in Eq.~(\ref{eq: positive edge energy}). This energy spectrum lies  inside the superconducting gap, and corresponds to edge states at the perimeter of the disc. One such edge state for $\ell = 0$ is shown in Fig.~\ref{fig: localized mode appen}. The edge mode we have found is the \textit{chiral} edge state we have sought from the beginning. From Eq.~(\ref{eq: positive edge energy}) we see that each of the Hamiltonians $\mathcal{H}_\pm$ describes
a single chiral band with unidirectional group velocity---negative for the case described here. This reflects the fact that a single chiral Hamiltonian breaks time-reversal invariance.
To recover time-reversal invariance, we now consider the full Hamiltonian, Eq.~(\ref{eq: fullhamiltonian}), of the two-dimensional topological superconductor. The spectrum of edge modes for $\mathcal{H}_-$ consists of the time-reversed states of $\mathcal{H}_+$. Suppose a state $\Psi_{+}^\text{edge}$ is an eigenstate of $\mathcal{H}_+$ of angular momentum $\ell$, then the time-reversed state with angular momentum $-(\ell+1)$ is an eigenstate of $\mathcal{H}_-$. Therefore, the edge spectrum of $\mathcal{H}_-$ is 
\begin{equation}
\label{eq: negative spectrum}
\mathcal{E}_{-}^{\text{edge}} \approx \frac{\ell+\frac{1}{2}}{\ell_0} \delta.
\end{equation}
This is just the chiral spectrum for modes propagating in the other direction. Taken together, Eqs.~\ref{eq: positive edge energy} and~\ref{eq: negative spectrum} give the full spectrum of the two edge modes for our time-reversal-invariant topological superconductor. The band structure of this system is shown in Fig.~\ref{fig: spectrum of edge states appen}.

\section{Majorana States in the Presence of a Zeeman Field}

\begin{figure}
\begin{center}
\includegraphics[scale=0.19]{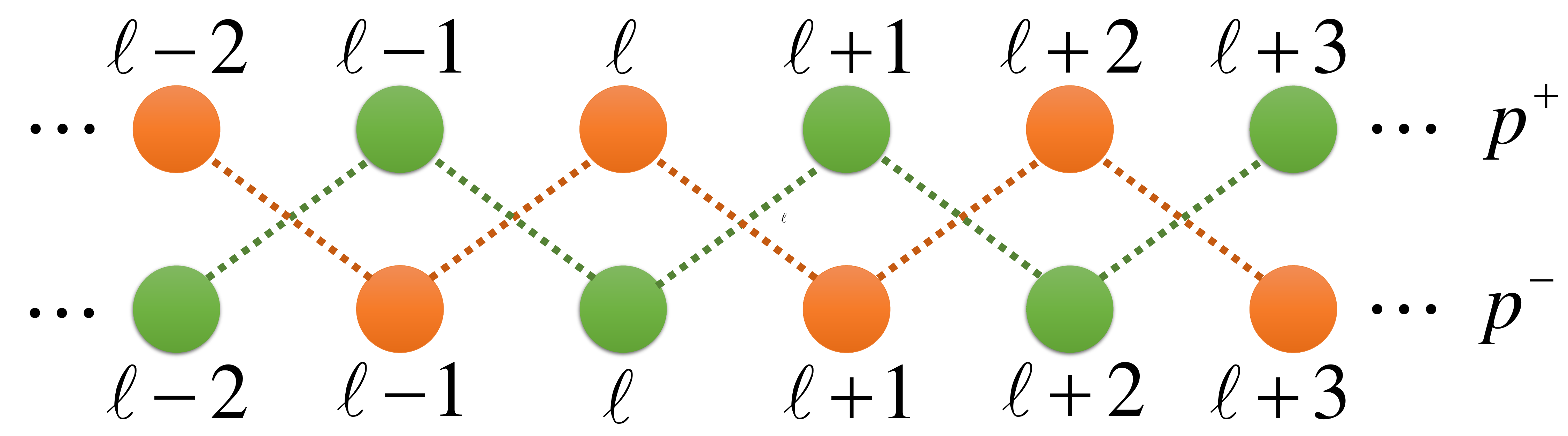}
\end{center}
\caption{Interaction between the two superconducting branches in the presence of an in-plane constant Zeeman field. We see that states with angular momentum $\ell$ from one branch couple to states with angular momenta $\ell \pm 1$ from the other branch. Therefore, the perturbation factorizes into two subspaces, resulting in a two-fold degeneracy in the zero-mode ground state.}
\label{fig: interaction ladder appen}
\end{figure}

\begin{figure}
\begin{center}
\includegraphics[width=8cm]{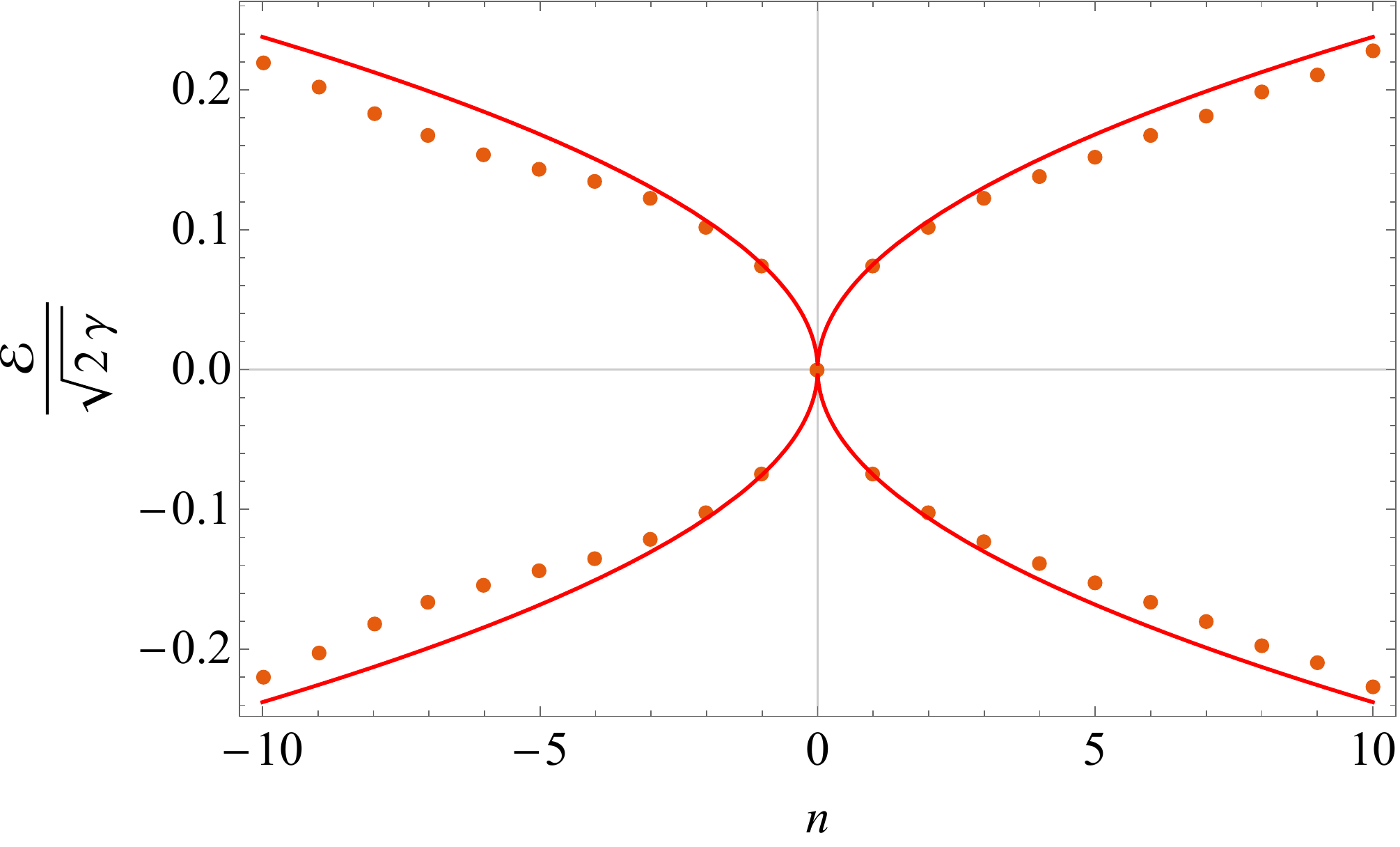}
\end{center}
\caption{Spectrum of the perturbed Hamiltonian. The dotted eigenvalues are obtained from diagonalizing the truncated Hamiltonian numerically. The solid lines are plotted using Eq.~(\ref{eq: edge energy in presence of field}). We use the following parameters: $\gamma = 1/16$, $\lambda = 50$, and $\mathcal{E}_Z = 1/20$. We label states by $n,$ which one can think of as the average angular momentum after mixing due to the perturbation. Note that the approximation in Eq.~(\ref{eq: edge energy in presence of field}) is in good agreement with the numerical eigenvalues for small $n$. 
}
\label{fig: perturbed spectrum appen}
\end{figure}

In this section, we consider the effect of a constant in-plane Zeeman field defined by 
\begin{equation}
\label{eq: Hamiltonian of perturbation}
\mathcal{H}_Z =  \tau_z(\mathcal{E}_Z\mathbf{n} \cdot \mathbf{ s}  ) = \mathcal{E}_Z \left( \begin{array}{cccc}
0 & 0 & 0 & e^{-i \phi}  \\
0 & 0 & - e^{i \phi}&0 \\
0&-e^{-i \phi}  & 0 & 0 \\
e^{i \phi} &0& 0 & 0   \end{array} \right),
\end{equation}
where $\mathcal{E}_Z$ is the energy magnitude of the field, $\mathbf{n}$ is a unit normal along the direction of the field, and $\phi$ is the angle of that field.
This Zeeman field couples electrons of opposite spins and holes of opposite spins, both with the same sign. By construction, this addition to the Hamiltonian explicitly breaks time-reversal symmetry. The full Hamiltonian now becomes
\begin{equation}
\mathcal{H} = \mathcal{H}_0 + \mathcal{H}_Z,
\end{equation}
where $\mathcal{H}_0$ is defined by Eq.~(\ref{eq: fullhamiltonian}). We  consider $\mathcal{H}_Z$ as a weak perturbation to $\mathcal{H}_0$, in comparison to the low-lying edge states with energies below the superconducting gap. That is, we assume $\mathcal{E}_Z\ll \delta$. So we can use lowest-order perturbation theory to assess how the energy spectrum of the low-lying edge states is changed. We first write explicitly the full edge wavefunctions. With $C_+$ and $C_-$ in Eq.~(\ref{eq: wavefunction})
 related by the vanishing boundary condition
\begin{equation}
 \frac{C_-}{C_+} =  \frac{\sqrt{2\gamma} \kappa_+ J_{\ell+1}(\lambda \kappa_+)}{\sqrt{2\gamma} \kappa_- J_{\ell+1}(\lambda \kappa_- )} = \frac{\left( \kappa_+^2-1-\mathcal{E}_+ \right)J_\ell(\lambda \kappa_+)}{\left( \kappa_-^2-1-\mathcal{E}_+ \right)J_\ell(\lambda \kappa_-)},
\end{equation}
we can write the wavefunction in the $\mathcal{H}_+$ subspace as 
\begin{equation}
\label{eq: positive wavefunction in symmetric form}
\begin{split}
|\ell,+\rangle = \mathcal{N} e^{i\ell\theta}&\left \{ \left( \begin{array}{ccc}
-  e^{i\theta} \left( \kappa_-^2-1-\mathcal{E}_+ \right) \kappa_+ J_\ell(\lambda\kappa_-) J_{\ell+1}(\lambda \kappa_+\rho) \\
\left( \kappa_+^2-1-\mathcal{E}_+ \right)\kappa_- J_{\ell+1}(\lambda \kappa_-)J_\ell(\lambda \kappa_+ \rho)\\
0\\
0\end{array} \right)-\left( \begin{array}{ccc}
- e^{i\theta}\left( \kappa_+^2-1-\mathcal{E}_+ \right)\kappa_- J_\ell(\lambda \kappa_+) J_{\ell+1}(\lambda \kappa_-\rho) \\
\left( \kappa_-^2-1-\mathcal{E}_+ \right) \kappa_+ J_{\ell+1}(\lambda \kappa_+)J_\ell(\lambda \kappa_-\rho)\\
0\\
0\end{array} \right) \right \},
\end{split}
\end{equation}
where $\mathcal{N}$ is a normalization constant.  We can simplify the wavefunction further by noting that
\begin{equation}
J^*_{\ell}(\lambda \kappa_+ \rho) = J_{\ell}(\lambda \kappa_+^* \rho) = J_{\ell}(\lambda \kappa_- \rho),
\end{equation}
where the last equality follows from inspection of Eq.~(\ref{eq: kappa}) that for $\mathcal{E} < \delta$, $\kappa_\pm$ is complex, and consequently, $\kappa_+^* = \kappa_-$. Using this, it is clear that the radial part of Eq.~(\ref{eq: positive wavefunction in symmetric form}) is purely imaginary
\begin{equation}
\label{eq: positive wavefunction in simplified form}
\begin{split}
|\ell,+\rangle = \mathcal{N} e^{i\ell\theta} \left( \begin{array}{ccc}
  e^{i\theta} \text{Im}\left[\left( \kappa_+^2-1-\mathcal{E}_+ \right)\kappa_- J_\ell(\lambda \kappa_+) J_{\ell+1}(\lambda \kappa_-\rho)\right] \\
\text{Im} \left[\left( \kappa_+^2-1-\mathcal{E}_+ \right)\kappa_- J_{\ell+1}(\lambda \kappa_-)J_\ell(\lambda \kappa_+ \rho)\right]\\
0\\
0\end{array} \right).
\end{split}
\end{equation}
The eigenstates of the $\mathcal{H}_-$ subspace are simply the time-reversed versions of the $\mathcal{H}_+$ states. We obtain the $\mathcal{H}_-$ eigenstates by taking the complex conjugate of Eq.~(\ref{eq: positive wavefunction in simplified form}), mapping $\ell \mapsto -(\ell+1)$, inverting the order of the components, and adding an appropriate minus sign
\begin{equation}
\label{eq: negative wavefunction}
\begin{split}
|\ell, -\rangle = \mathcal{N} e^{i\ell\theta} \left( \begin{array}{ccc}
 0\\
 0\\
-e^{i\theta} \text{Im} \left[\left( \kappa_+^2-1-\mathcal{E}_- \right)\kappa_- J_{\ell}(\lambda \kappa_-)J_{\ell+1}(\lambda \kappa_+ \rho)\right]\\
  \text{Im}\left[\left( \kappa_+^2-1-\mathcal{E}_- \right)\kappa_- J_{\ell+1}(\lambda \kappa_+) J_{\ell}(\lambda \kappa_-\rho)\right] \\
\end{array} \right).
\end{split}
\end{equation} 
In this form, we have electron and hole components in the order defined in Eq.~(\ref{eq:psiorder}).
We can now  study the coupled Hamiltonian by diagonalizing $\mathcal{H}_Z$ in this basis. There is a natural cut-off on the value of $\ell$ using the fact that the square-root in Eq.~(\ref{eq: kappa}) must be real
\begin{equation}
|\ell|<\ell_\text{max}=\lambda \sqrt{1-2\gamma}.\label{eq:ellmax}
\end{equation}
Here, we see that $0<\gamma < 2.$ However, as we will show in a later section, it is, in fact, the case that for a range of semi-realistic parameters, $\gamma \ll 1.$  The diagonal elements of the matrix then contain the unperturbed energies, with the upper-left  block containing the $p^+$ spectrum, and the lower-right block containing the $p^-$ spectrum.
The Zeeman interaction only mixes electrons and holes of opposite spins, and thus couples the $p^+$ and $p^-$ subspaces. The matrix elements are of the form
\begin{equation}
\langle \ell+m,-| \mathcal{H}_Z |\ell,+\rangle = f_\text{rad} \int_0^{2\pi} e^{-i(m+1)\theta}  d\theta + g_\text{rad} \int_0^{2\pi} e^{i(m-1)\theta}  d\theta \propto \delta_{m,\lbrace  \pm 1 \rbrace},\label{eq:Hzme}
\end{equation} 
where $f_\text{rad}$ and $g_\text{rad}$ are radial integrals.  
Therefore, we see that this perturbation only couples states of angular momenta $\ell$ and $\ell\pm 1$. This is shown schematically in Fig.~\ref{fig: interaction ladder appen}. The Hamiltonian can now be diagonalized numerically to find the spectrum of edge modes. Since angular momentum is not conserved because the perturbation is not axis-symmetric, we label states of the new spectrum by $n$ instead of $\ell$. One can think of $n$ as the average angular momentum after hybridization from the perturbation. The result of a particular realization of this perturbed model is shown in Fig.~\ref{fig: perturbed spectrum appen}. In this case, we use $\lambda = 50$ and $\mathcal{E}_Z = 1/20$. As shown in Fig.~\ref{fig: perturbed spectrum appen}, adding this perturbation generates a gap, but there remain two localized modes at almost zero energy. This is due to particle-hole symmetry that is preserved by the perturbation. In other words, for every positive-energy state, there exists a negative-energy state of the same amplitude in the unperturbed spectrum. The addition of  the Zeeman perturbation shifts the energies, but preserves the fact that there is a hole state for every particle state. Hence, the states with approximately zero energy come in pairs. Furthermore, we find numerically that the radial integrals in Eq.~(\ref{eq:Hzme}) are almost exactly 1/2, with small deviations only when $\ell$ gets close to $\pm\ell_\text{max}$.

We now estimate the magnitude of resulting gap due to the perturbation. As we have established, the spectra of the unperturbed edge states can be well-approximated by a linear energy dispersion of $\ell+1/2$ with opposite chirality. Therefore, we can write the Hamiltonian corresponding to this two-state system as 
\begin{equation}
\label{eq: Hamiltonian for edge states}
\mathcal{H}_0^\text{edge} = \left( \begin{array}{cccc}
 \frac{\sqrt{2\gamma}}{\lambda}\left( \ell + \frac{1}{2} \right) & 0  \\
0 & -\frac{\sqrt{2\gamma}}{\lambda}\left(\ell + \frac{1}{2} \right)  \end{array} \right).
\end{equation}
The Zeeman field couples states in one branch with angular momentum $\ell$ with states in the other branch with angular momentum $\ell \pm 1$. We can correspondingly decompose the wavefunctions as
\begin{align}
\left| \Psi \right\rangle &= \sum_\ell a_{\ell , +} | \ell, + \rangle + \sum_\ell a_{\ell , -} | \ell, - \rangle,
\end{align}
where the quantities $a_{\ell , \pm}$ are amplitudes. Next, we use
 ${\cal H}_Z | \ell, + \rangle  = \frac{\mathcal{E}_z}{2} ( | \ell+1,- \rangle  + |\ell - 1 , -\rangle )$. We can make a gauge transformation which changes the sign of one half of the amplitudes $a_{\ell ,\pm}$, which is nothing more than a $\pi/2$ rotation of the axes. Thus, if these new amplitudes are slowly varying,  we find that for the matrix representation of ${\cal H}_Z$, we can write
${\cal H}_Z a_{\ell , +} \approx \mathcal{E}_z \frac{\partial a_{\ell , -}}{\partial \ell}$. The  Hamiltonian can now be approximately expressed as
\begin{align}
{\cal H}^\text{edge} &= \frac{\sqrt{2 \gamma}}{\lambda} \left( \ell + \frac{1}{2} \right) \sigma_z +  i \mathcal{E}_Z \sigma_x \partial_\ell = \frac{\sqrt{2 \gamma}}{\lambda} \left( \hat{\ell} + \frac{1}{2} \right) \sigma_z -  \mathcal{E}_Z \sigma_x \hat{ p}_\ell,\label{eq:Hedge}
\end{align}
where $\sigma_x$ and $\sigma_z$ are Pauli matrices, and we define $\hat{\ell}$ and $\hat{ p}_\ell$ as the position and momentum operators. By squaring the Hamiltonian, we find
\begin{align}
\left( {\cal H^\text{edge}} \right)^2 &= \left(\frac{2 \gamma}{\lambda^2} \left( \hat{\ell} + \frac{1}{2} \right)^2 +  \mathcal{E}_Z^2 \hat{ p}_\ell^2\right)\mathcal{I} - \frac{ \mathcal{E}_Z \sqrt{2 \gamma}}{\lambda} \sigma_y .
\end{align}
This squared Hamiltonian describes a harmonic oscillator. The eigenenergies are, using $m$
as the oscillator quantum number,
\begin{align}
(\mathcal{E}_\pm^\text{edge})^2 &= \frac{2 \mathcal{E}_Z \sqrt{2 \gamma}}{\lambda} \left( m + \frac{1}{2} \right) \pm \frac{ \mathcal{E}_Z \sqrt{2 \gamma}}{\lambda}.
\end{align}
This shows that the zero-energy mode occurs once, and every other energy 
occurs twice. We can thus express the final spectrum of the edge Majorana modes in terms of $n=\pm m$ as
\begin{align}
\mathcal{E}_\pm^\text{edge} &= \pm \sqrt{ \frac{2 \mathcal{E}_Z \sqrt{2 \gamma}}{\lambda} | n |},\label{eq: edge energy in presence of field}
\end{align}
where $n =..., -2, -1, 0 , 1 , 2 , .$.. The wavefunction of the zero-energy Majorana modes is a Gaussian in angular momentum space with harmonic oscillator length parameter 
\begin{equation}
\label{eq: angular width inverse}
b_\ell=\left(\frac{(\mathcal{E}_Z\lambda)^2}{2\gamma}\right)^{\frac{1}{4}}.
\end{equation}
This parameter is $\gtrsim 1$, which agrees with the fact that we have assumed a slow variation of the coefficients $a_{\ell,\pm}$. The implication is that the Gaussian width in the dual, and more physically relevant, variable $\theta$ is given by $1/b_{\ell}$ and is reasonably narrow. This implies that the wavefunction of the Majorana zero-modes is narrowly localized in angular space.  We can see that this width decreases as we increase the radius of the disc or the coupling to the magnetic field.

The expression (\ref{eq: edge energy in presence of field}) is valid when the width in $\theta$ is so small that the Majorana wavefunctions do not significantly overlap. As can be seen in Fig.~\ref{fig: perturbed spectrum appen}, the harmonic spectrum fits well for realistic parameters, but works best for small $n$, since the approximation of a very narrow state is not completely satisfied. We can improve the agreement by increasing the value of $\lambda$. From these results, we can estimate that the gap between the Majorana zero-modes and the nearest quasiparticle states is given by
\begin{equation}
\Delta_{M}=\sqrt{ \frac{2 \mathcal{E}_Z \sqrt{2 \gamma}}{\lambda} },\label{eq:DeltaM}
\end{equation}
which is larger than the spacing between the first quasiparticle states by a factor of 
$1/(\sqrt{2}-1)\approx 2.4$.

The derivation above assumes three simplifications, which are all irrelevant for the
estimate of the quasi-particle gap $\Delta_M$, but have consequences for the energies of the mid-gap states that are slightly displaced from zero. We thus have a gap in the spectrum of the quasi-zero-modes, which we shall denote by $\delta\epsilon_M$. This effect that the Majorana states are actually only quasi-Majorana states is common to all approaches to topological computing.

The three approximations are, in order of the severity of their impact on $\delta\epsilon_M$:
\begin{enumerate}
\item The approximation that there is no upper limit for $\ell$, whereas we know that $\ell_\text{max}$ is finite (see Eq.~(\ref{eq:ellmax})),
\item The assumption that the radial matrix elements in Eq.~(\ref{eq:Hzme}) are all $1/2$, and
\item The neglect of the periodicity of $\theta$  in Eq.~(\ref{eq:Hedge}).
\end{enumerate}
We now investigate all of these here. The first point can be dealt with exactly by solving the harmonic oscillator problem with a cut-off on $\ell$ by placing our harmonic oscillator into an infinite well in $\ell$ space with hard walls located at $\pm \ell_\text{max}.$ This will only make a significant contribution when the wavefunction in $\ell$ is wide compared to $\ell_\text{max}$, and is proportional to $\exp(-b_\ell^2/\ell_{\text{max}}^2)$, which is extremely small for a very wide range of parameters. Therefore, this contribution to the splitting of the mid-gap energies is not relevant for the system considered here. Since there is so little sensitivity to the behavior at large values of $\ell$, the second point is also not important; we find that
the dominant source of splitting comes from the periodicity in $\theta$.

It is straightforward to show that, after making the same change-of-sign transformation on the coefficient $a_\ell$ as before, but not ignoring the fact that $\theta$ is periodic, we find that the Hamiltonian (\ref{eq:Hedge}) including periodicity is
\begin{align}
{\cal H}^\text{edge} &= \frac{\sqrt{2 \gamma}}{\lambda} \left( i\partial_\theta + \frac{1}{2} \right) \sigma_z - \mathcal{E}_Z \sigma_x \sin2\theta.\label{eq:Hedge2}
\end{align}
Once again we can square this Hamiltonian, but the result is not as illuminating as before,
\begin{align}
\left( {\cal H^\text{edge}} \right)^2 &= \left(\frac{2 \gamma}{\lambda^2} \left( i\partial_\theta  + \frac{1}{2} \right)^2 +  \mathcal{E}_Z^2 \sin^22\theta\right)\mathcal{I} - \frac{ \mathcal{E}_Z \sqrt{2 \gamma}}{\lambda}\cos2\theta \,\sigma_y,
\end{align}
which is a set of two Mathieu-like equations with shifted derivative with a coupling term.

We have not attempted to solve this problem in a general way, even though there are many indications
that this may be possible. Instead, we use a combination of numerical calculations with some analytical insights to find an appropriate dependence of the splitting of the mid-gap energies on this periodicity in $\theta.$ We start from Eq.~(\ref{eq:Hedge2}), and perform a scaling inspired by the harmonic oscillator results (for $\Delta_M$ see Eq.~(\ref{eq:DeltaM}))
\begin{align}
{\cal H}^\text{edge} &= \Delta_M \left(\frac{1}{b_\ell}\left( i\partial_\theta + \frac{1}{2} \right) \sigma_z - b_\ell \sigma_x \sin2\theta\right).
\end{align}
Thus we know that all eigenenergies can be exactly be expressed in the form $\Delta_M f_n(b_\ell),$ where $f_n$ is some function that depends on the energy level $n$ and the parameter $b_\ell.$
Indeed,  numerical diagonalizations  with a finite cutoff $\ell_\text{max}$ give exactly this scaling behavior. 
Actually, in the case of small splitting, we find that a good approximation for $f_0$ is given by 
\begin{equation}
f_0(b_\ell)=\exp(-2 b_\ell^2)/\sqrt{2},
\end{equation}
with no deviations until $f_0$ gets as large as $0.1$--well outside the regime of interest. More typical values of $b_\ell=2-3$ give $f_0=2\times10^{-4}-10^{-8}$. This establishes that the splitting of the mid-gap modes is very small.

\section{Estimates of Physical Parameters}

In this section, we give estimates of the physical parameters to hopefully guide future experimental attempts to realize our proposal. We note that although our estimates will based primarily on data that are available on Cu$_x$Bi$_2$Se$_3,$ our proposal is not limited to just this two-dimensional material, as we have already explained in the main text.

First, we estimate the physical length scale of the disc. The three-dimensional electron density of Cu$_x$Bi$_2$Se$_3$ has been reported to be about $n \approx  2 \times 10^{20}$ cm$^{-3}$ \cite{Hetal10}. This value varies slightly depending on dopant concentration of Cu \cite{PhysRevB.84.054513}. Using the fact that each quintuple layer of  Cu$_x$Bi$_2$Se$_3$ is about $1$ nm \cite{doi:10.1063/1.3200237}, we find that the electron density per layer is $\rho_L \approx 1.6 \times 10^{13}$ cm$^{-2}.$ The Fermi wavelength is then given by $k_F = \sqrt{2\pi \rho_L} = 10^{-1}$ \AA$^{-1}.$ We want the Fermi wavelength to be much smaller than the radius of the disc. Therefore, we choose $\lambda$ to be large. For instance, with $\lambda = 100,$ the corresponding radius is $R = 1$ $\mu$m. For the simulations shown in this paper, we have chosen $\lambda = 50$ to improve visibility as the edge modes become too localized to clearly see with increasing radius.

Next, we determine the various energy scales of the proposal. The critical temperature $T_c$ is found experimentally to be 3.8 K \cite{Hetal10}. The superconducting gap is then given by $\Delta_\text{exp}=1.76 k_BT_c =k_B \times 6.7 \, \mathrm{K} = 0.6 \, \mathrm{meV}$ . We note that scalar disorder can break $p$-wave superconducting pairs, leading to a lower $T_c.$ This means that it is possible in future experiments that cleaner samples can host superconductivity with a higher $T_c$ and larger gap. It has been reported that the effective mass $m_\text{eff}$ in Cu$_x$Bi$_2$Se$_3$ is about 0.3 $m_e$ electron masses \cite{PhysRevB.90.094503}. Using this, we can estimate the Fermi energy $\epsilon_F = \hbar^2 (3\pi^2 n)^{2/3}/2m_\text{eff} \approx 350 $ meV. Now, if we take $E_G = 150$ meV \cite{PhysRevB.83.224516}, this then gives the shifted Fermi energy $\bar{\epsilon}_F = \epsilon_F-E_G = v_F^2k_F^2/2E_G = 200 \, \mathrm{meV} = k_B \times 2300 \, \mathrm{K}.$ The coherence length in the BCS theory is given by $\xi = \hbar v_F/ \pi \Delta_\text{exp} = \bar{\epsilon}_F/\Delta_\text{exp} k_F \approx 3.4 \times 10^3$ \AA. This calculation justifies our assumption of quasi-two-dimensionality because the coherence length is much larger than the thickness of each quintuple layer of Cu$_x$Bi$_2$Se$_3$.

Finally, we estimate the physical quantities relevant when a Zeeman field $B_Z = 1$ T is applied. When a magnetic field is applied, Majorana zero-modes emerge, localized in angular space where the field is tangent to the disc's surface. The angular width of the Majorana states is given by the inverse of Eq.~(\ref{eq: angular width inverse}), $b_\ell^{-1} = \sqrt{\Delta_\text{exp}/\mu_e B_Z Rk_F} \approx 0.32,$ where $\mu_e$ is the Bohr magneton of the electron. The energy gap between the zero-modes and the first quasiparticle excitations is $\Delta_M  = \sqrt{2\mu_eB_Z \Delta_\text{exp}/Rk_F} \approx 0.026$ meV $ = k_B\times 0.3$ K. The gap between the quasi-zero-energy Majorana modes is $\delta \epsilon_M = \Delta_M \exp(-2b^2)/\sqrt{2} \approx  10^{-9} $ meV $ = k_B \times 10^{-8}$ K.

We note that the above parameters are semi-realistic estimates. Using these, the dimensionless paramters are $\lambda = 100,$ $\gamma < 10^{-5}$ and $\mathcal{E}_Z < 10^{-3}.$ However, in numerical simulations, we have chosen these dimensionless parameters to be $
\lambda = 50, $ $\gamma = 1/16,$ and $\mathcal{E}_Z = 1/20$ for improved visibility. 

\section{Tight-Binding Calculations}
\begin{figure}
\begin{center}
\includegraphics[width=5cm]{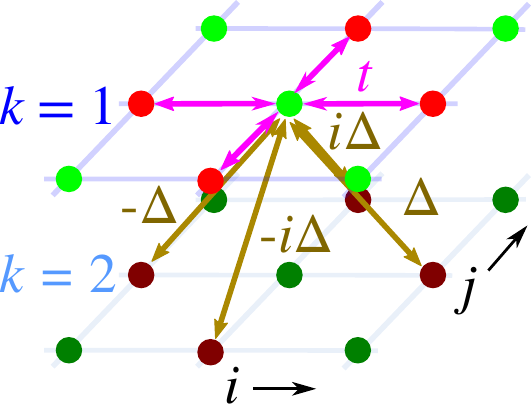}
\end{center}
\caption{A schematic representation of the tight-binding
model for a single spin projection (i.e. the $p_x+ip_y$ model) on a square lattice.}
\label{fig:tight1 appen}
\end{figure}

\begin{figure}
\includegraphics[width=10cm]{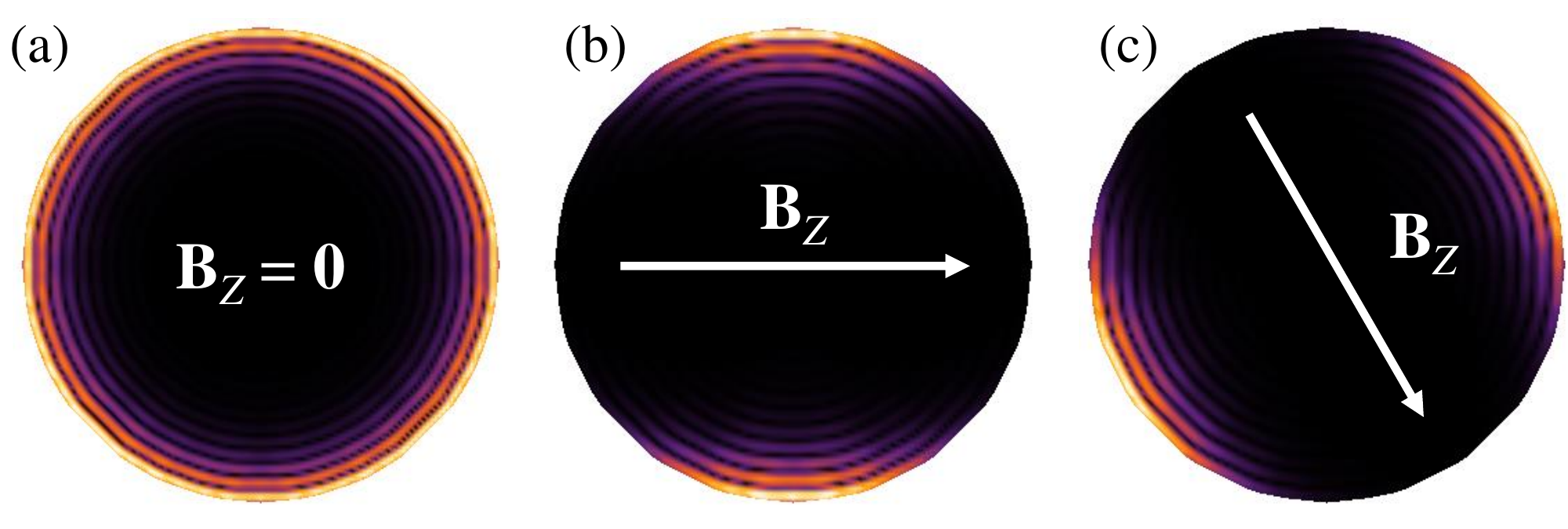}
\caption{Electron density for the tight-binding model on a disc-like geometry. In each case,
we only show one of the four components of the corresponding spinor wavefunction as all four components have the same probability density.
(a) Edge state for a single chiral Hamiltonian in the absence of an in-plane Zeeman field. 
(b) Majorana state for a Zeeman field in the $x$ direction and (c) Majorana state for a field 
making an angle $\phi = \pi/3$ with the $x$-axis. The radius of the circle is 50 lattice spacings.
\label{fig:TBcirc appen}}
\end{figure}

In this section, we present alternative calculations of our model using the tight-binding approach. This can either be thought of as
a microscopic realization of the continuum model or a
finite-difference discretization. We choose to work on a
square lattice. However, the results obtained for system sizes much larger than
the lattice spacing should be independent of this choice. The process of nearest-neighbor hopping
is most easily described on a bipartite lattice. For a single chiral superconductor, corresponding to a single spin projection, the gap couples the conduction band to the valence band. Each occupied site has an
energy $\pm \epsilon_F$ (plus for particles, minus for holes), and the hopping
constant is $\pm t$ (where the sign indicates the direction of hopping). The superconducting gap $\Delta_\text{sc}$ couples particles to holes with the usual Dirac matrix structure, as
shown in Fig. \ref{fig:tight1 appen}.
The full model then combines the spin-up and spin-down copies of the chiral superconductor with
conjugate Hamiltonians. The Zeeman term couples particles with spin up and spin down, and similarly for the
holes. In summary, if $\mathbf{r}=(i, j)$ labels the sites, $k = 1, 2$ labels
particles or holes respectively, $l = 1, 2$ labels spin up or
down respectively,  $\boldsymbol{\rho}$ links nearest-neighbor points, $a$ is the lattice spacing, $z$ is the coordination number, and the indices on the Pauli matrices denote a matrix element, then the Hamiltonian has the following form on any bipartite lattice:
\begin{align}
\mathcal{H}_{\mathbf{r}kl;\mathbf{r}'k'l'}&= 
s_{z,ll'} \tau_{z,kk'}  \left[ -\delta_{\mathbf{r}\mathbf{r}'}\mu
- \frac{1}{2}\sum_{\boldsymbol \rho}\delta_{\mathbf{r},\mathbf{r}'+\boldsymbol \rho} t\right]
+\Delta_\text{sc} \delta_{ll'}\frac{i}{2}\sum_{\boldsymbol \rho}\delta_{\mathbf{r},\mathbf{r}'+\boldsymbol{\rho}}
[{\tau}_{x,kk'}{\rho_y}+{\tau}_{y,kk'}{\rho_x}]
+\mathcal{E}_Z \delta_{\mathbf{r}\mathbf{r}'}\tau_{z,kk'}\mathbf{n} \cdot \mathbf{s}_{ll'}.
\end{align}
This has the continuum limit
\begin{align}
\mathcal{\hat{H}}_{kl;k'l'}&=s_{z,ll'} \tau_{z,kk'}  \left[ -\mu+\frac{z}{2}t 
-\frac{z a^2t }{8} \hat{p}^2\right]
+\frac{z a^2}{4}\Delta_\text{sc} \delta_{ll'}
[{\tau}_{x,kk'}{\hat{p}_y}+{\tau}_{y,kk'}{\hat{p}_x}]
+ \mathcal{E}_Z \tau_{z,kk'} \mathbf{n} \cdot \mathbf{s}_{ll'},
\end{align}
which makes it simple to read off the relation between the parameters of the continuum and tight-binding models. With this tight-binding model, we can study any geometry, and on any bipartite lattice contained
within that geometry. For the results reported here, we use a square lattice, but we have 
checked that we can obtain the same answers for a hexagonal lattice.
We first study  a disc, or rather, an approximation to a disc built from squares, to compare with our analytic calculations on the same geometry. We find that the numerical results are identical to those of the continuum model, as shown in Fig.~\ref{fig:TBcirc appen}a. What cannot be seen from that figure is the angular dependence of the resulting wavefunctions; these are in agreement with the continuum model as well. If we add an in-plane magnetic field,
we find the Majorana zero-modes as seen before, shown in Fig.~\ref{fig:TBcirc appen}b-c. We note that even if the field is not aligned along a symmetry axis of the lattice, where the lattice has more effect, our tight-binding model still produces consistent results, as illustrated in Fig.~\ref{fig:TBcirc appen}c. 

\begin{figure}
\includegraphics[width=17.5cm]{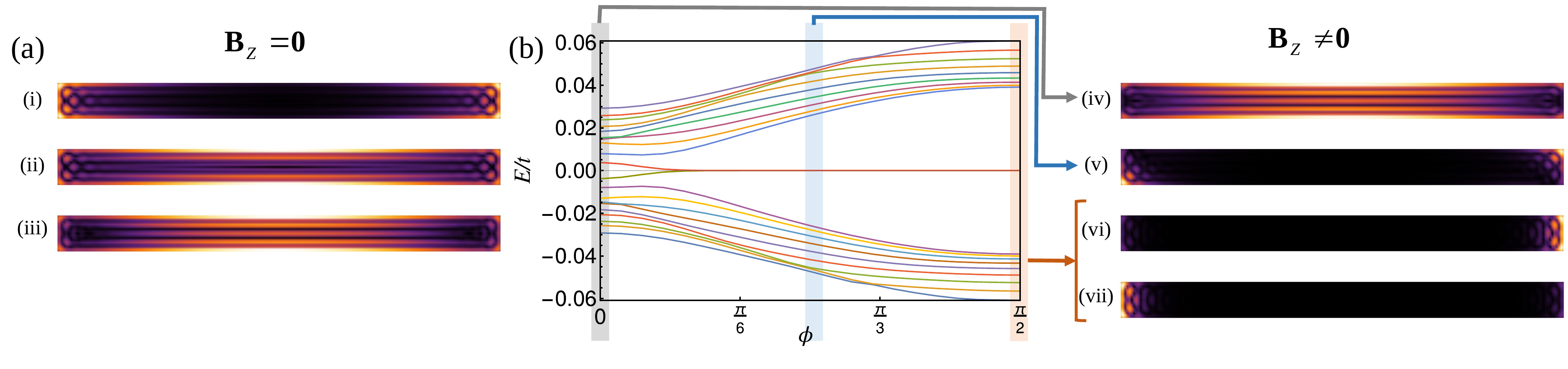}
\caption{Tight-binding model on a 2D wire. The dimensions of the wire are $100 \times 8$ lattice units. (a) Electron densities for the tight-binding $p_x + i p_y$ model in the absence of an in-plane Zeeman field. In each case, we only show the electron probability. (i) The lowest absolute energy state; there is another state with the opposite energy of the same magnitude. (ii)-(iii) States at the next highest energy in absolute value with (ii) showing the positive-energy state and (iii) showing the negative-energy state. (b) Energy spectrum of the tight-binding model in the presence of a Zeeman field as a function of the field direction and electron densities corresponding to this setup. An angle of zero corresponds to a field parallel to the $x$-axis. We use $\mu = -3.2 ,$ $\Delta_\text{sc} = 0.2 t,$ and $\mathcal{E}_Z = 0.05 t.$  (iv)-(vii) Electron densities at different angles of the Zeeman field. We see that at angles $\phi \approx 0,$ there are no zero-modes, and correspondingly, (iv) shows that the lowest-energy state is a delocalized state. (v) Majorana state when the Zeeman field makes a $\pi/4$ angle with the $x$-axis. (vi)-(vii) Majorana states when the Zeeman field are parallel to the $y$-axis.}
\label{fig:wire appen}
\end{figure}

Next, due to our choice of a square lattice, the simplest geometry to study is that of a 2D quantum wire, the 
de-facto standard for topological quantum computing. We look for states with energies below the superconducting gap. In the case of a disc, all of these states are boundary states localized near the edge. However, in a wire geometry, we no longer have this property as some of these mid-gap states behave more like bulk states. As can be seen in Fig.~\ref{fig:wire appen}i-ii, in the absence of a Zeeman field, we find
 states in the $p_x+ip_y$ superconductor that are localized at the ends of the wire in the two lowest-energy states.  However, for the next  state higher in energy, the mid-gap state looks more like a bulk state, as shown in Fig.~\ref{fig:wire appen}iii. When we turn on the in-plane magnetic field perpendicular to the long axis of the wire, we get two Majorana states, each located at one end of the wire, as shown in Fig.~\ref{fig:wire appen}vi-vii. The numerical results violate reflection symmetry, due to an exact degeneracy in the spectrum. As we rotate the field, we see that this degeneracy is broken, and we move from states with the Majorana modes located at the ends to delocalized states as the field aligns with the long axis. This can be seen most clearly in the energy-level plot in Fig.~\ref{fig:wire appen}, where it is shown how the degeneracy in zero-energy states is lifted as we get close to the case where the field is parallel to the wire direction.

\begin{figure}
\includegraphics[width=9cm]{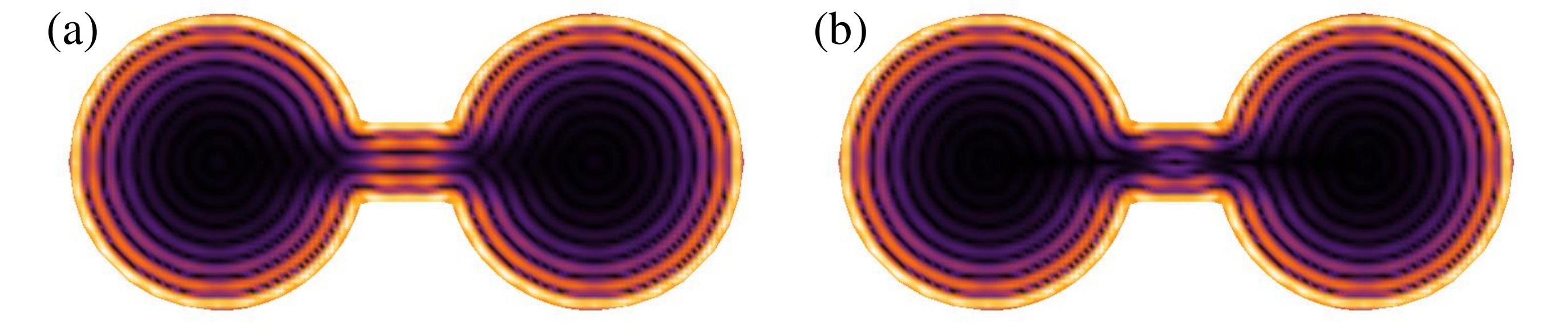}
\caption{Electron density (a) and hole density (b) for the $p_x+ip_y$ tight-binding model on two discs connected by a short wire. Each disc has a radius of $30$ lattice spacings.
We note the difference in the probability density in the connecting wire that requires further investigation.}
\label{fig:dumbbellnofield appen}
\end{figure}

\begin{figure}
\includegraphics[width=12cm]{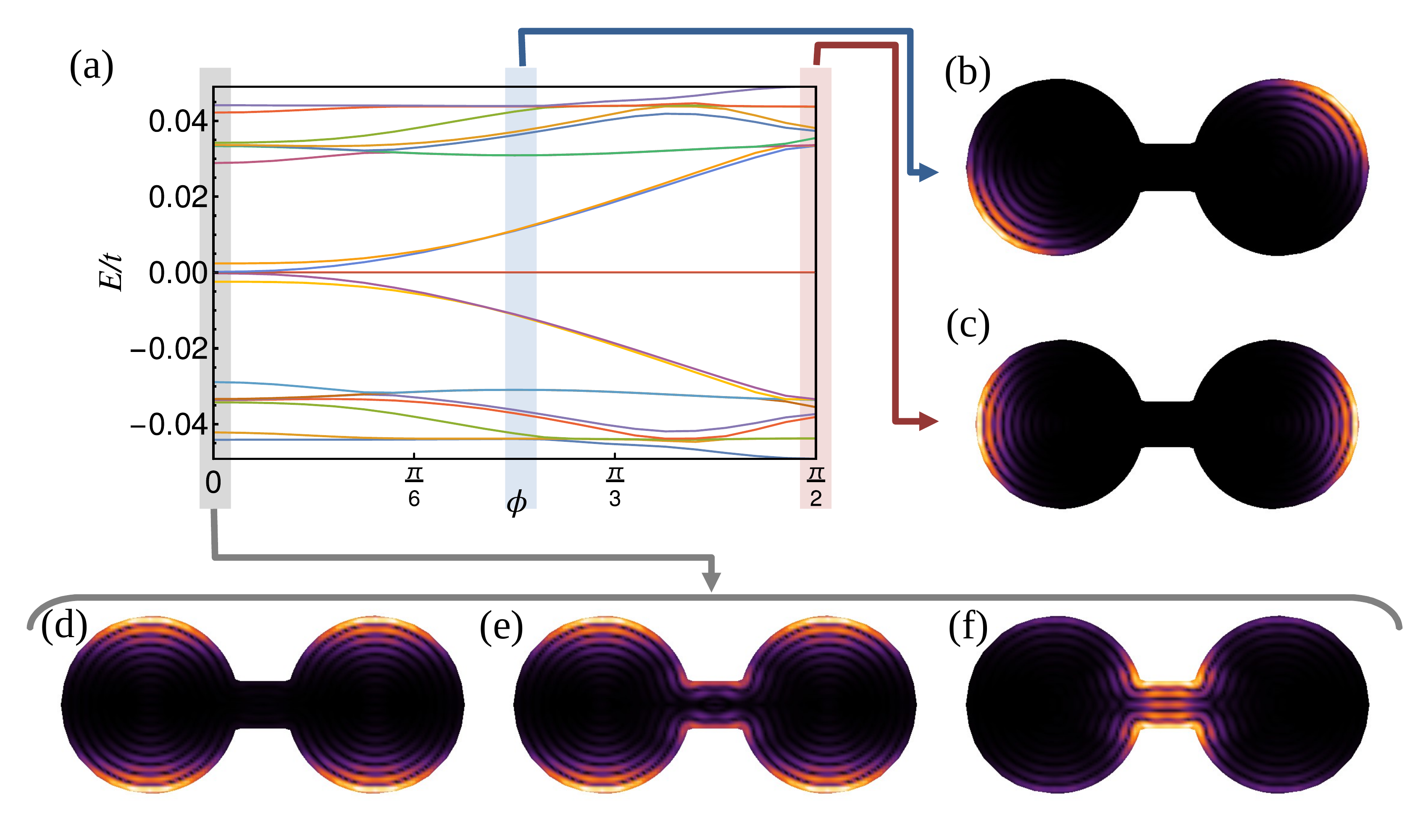}
\caption{Energy spectrum (a) for a $p_x+p_y$ superconductor in a dumbbell geometry with a thick connecting wire in the presence of a Zeeman field as a function of angle and electron density plots (b)-(f) for various angles of the field. Each disc of the dumbell has a radius of 30 lattice units. The connecting wire has length 16 and width 8 lattice spacings. We use $\mu=-3.2 t$, $\Delta_\text{sc}=0.2t$ and $\mathcal{E}_Z=0.05t$. In each of the density plots, we only show one of the four components of the spinor wavefunctions as the others behave similarly. (b) Majorana state when the Zeeman field makes a $\pi/4$ angle with the $x$-axis. (c) Majorana state when the Zeeman field is parallel to the $y$-axis. (d)-(f) The three Majorana-like states for a field parallel to the $x$-axis, in order of energy.}
\label{fig:thickwire appen}
\end{figure}

Of more interest to quantum-information processing is entangling the  Majorana states
in two adjacent discs connected by thin wires by rotating the magnetic field. Again, we first look at the $p_x+i p_y$ model in this setup in the absence of a Zeeman field, as shown in Fig.~\ref{fig:dumbbellnofield appen}. We see that inside the discs,
there is no distinction between the behavior of electrons and holes, but in the connecting wire, there is an interesting difference. This has not yet been investigated in detail in this work, but will likely have an impact on how to construct analytic models of this situation.
We now move onto assessing the model in the presence of a Zeeman field. For a  field that is not parallel to the connecting wire,  we find the expected Majorana states, as seen in Fig.~\ref{fig:thickwire appen}b-c. As we turn on the field parallel to the wire, we find that more states come into play, as shown in Fig.~\ref{fig:thickwire appen}d-f. We see  very low energy states that
are completely located only in the connecting wire, which is clearly a large perturbation on the situation of two unconnected discs.

\begin{figure}
\includegraphics[width=12cm]{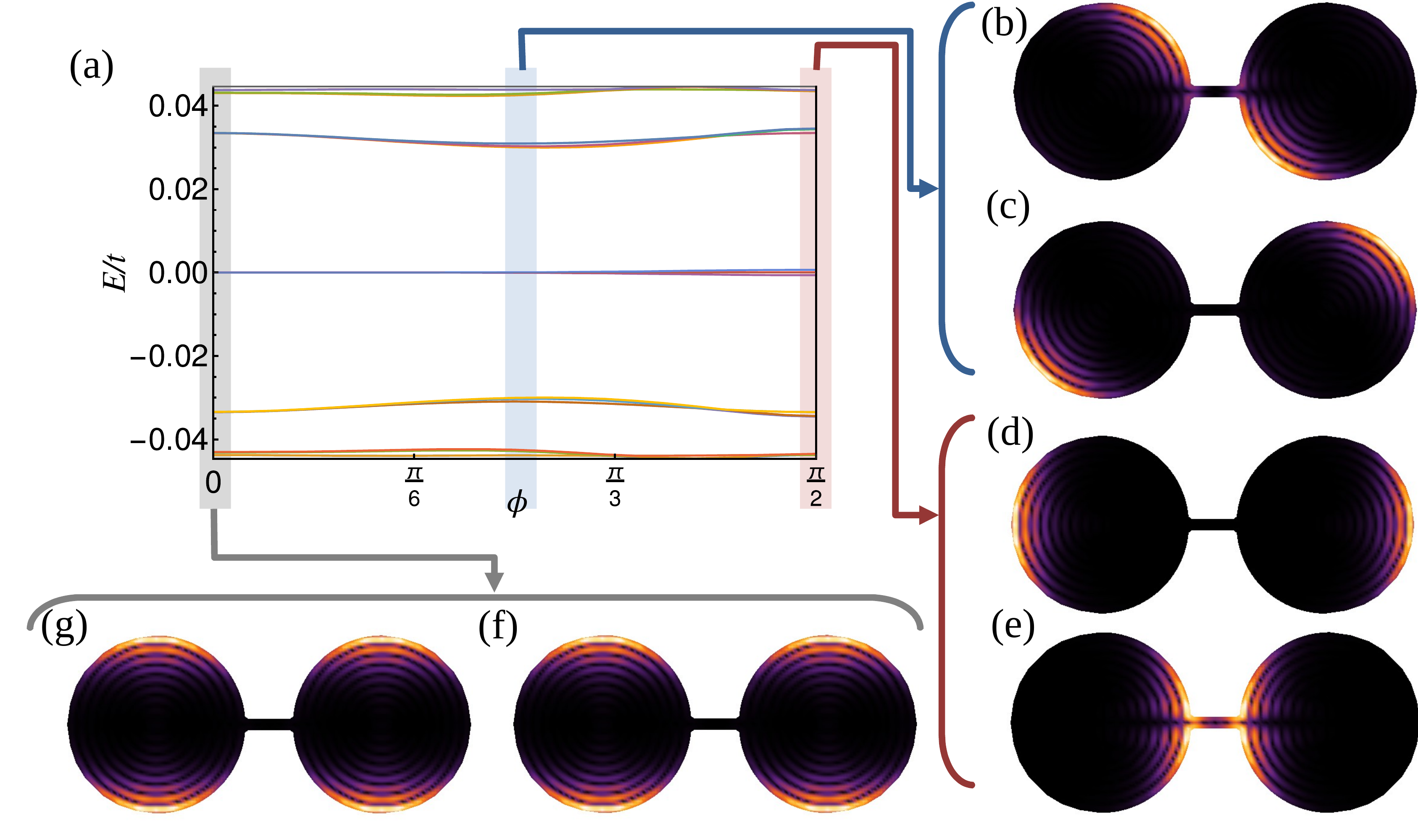}
\caption{Energy spectrum (a) for a $p_x+p_y$ superconductor in a dumbbell geometry with a thin connecting wire in the presence of a Zeeman field as a function of angle and electron density plots (b)-(g) for various angles of the field. Each disc of the dumbell has a radius of 30 lattice units. The connecting wire has length 16 and width 4 lattice spacings. We use $\mu=-3.2 t$, $\Delta_\text{sc}=0.2t$ and $\mathcal{E}_Z=0.05t$. In each of the density plots, we only show one of the four components of the spinor wavefunctions as the others behave similarly. (b)-(c) The two Majorana states for a Zeeman field makes a $\pi/4$ angle to the $x$-axis. (d)-(e) The two Majorana states when the Zeeman field is parallel to the $y$-axis. (f)-(g) The two Majorana-like states for a field parallel to the $x$-axis.}
\label{fig:thinwire appen}
\end{figure}

Since we want to reduce the coupling between the two superconducting discs, we consider a long, thin connecting wire next, with results shown in Fig.~\ref{fig:thinwire appen}. We see that in such a situation, we can instead realize a weak
perturbation, but there is still non-zero probability of the zero-modes penetrating the wire, and thus
the Majorana states still couple.


A most straightforward way to assess the coupling quantitatively is to look at the energy spectra of the two setups of either a thick or a thin connecting wire. Comparing Fig.~\ref{fig:thickwire appen}a and Fig.~\ref{fig:thinwire appen}a, we see that for a thicker wire, we get a strong mixing between quasiparticles and Majorana states, i.e. a clear example of quasiparticle poisoning. For a much thinner wire,  we get
almost no coupling between the Majoranas. The tunneling splitting is much smaller than the gap
to the first quasiparticle state. The latter case of a thinner wire would thus make a better candidate for connecting the three-dot device discussed in the main text for the purposes of performing quantum logic.

\end{document}